\documentclass[english]{emulateapj}
\usepackage{epsfig,amsmath,natbib}
\usepackage[bookmarks=false]{hyperref}
\usepackage{url}
\usepackage{multirow}
\usepackage{longtable,subfigure,float}
\newcommand{\todo}{\ifmmode \text{\Huge{\(\bullet\)}} \else {\Huge$\bullet$}\fi}
\newcommand{\tido}{\ifmmode {\bullet} \else $\bullet$\fi}

\newcommand{\E        }[1]{\ifmmode 10^{#1} \else $10^{#1}$\fi}
\newcommand{\tE        }[1]{\ifmmode \times10^{#1} \else $\times10^{#1}$\fi}
\newcommand{\til}{\ifmmode \sim \else $\sim$\fi}
\renewcommand{\~} {\ifmmode \sim \else $\sim$\fi}

\newcommand{\pc}	{\ifmmode {\rm pc} \else pc\fi}
\newcommand{\ld}	{\ifmmode {\rm l.d.} \else l.d.\fi}
\newcommand{\kms}	{\ifmmode {\rm km\,s}^{-1} \else km\,s$^{-1}$\fi}
\newcommand{\cc}	{\ifmmode {\rm cm}^{-3}    \else cm$^{-3}$\fi}
\newcommand{\cmii}	{\ifmmode {\rm cm}^{-2}    \else cm$^{-2}$\fi}
\newcommand{\ergs}	{\ifmmode {\rm erg\,s}^{-1} \else erg s$^{-1}$\fi}
\newcommand{\ergcms}	{\ifmmode {\rm erg\,cm}^{-2}\,{\rm s}^{-1} \else erg\,cm$^{-2}$\,s$^{-1}$\fi}
\newcommand{\ergcmsA}	{\ifmmode {\rm erg\,cm}^{-2}\,{\rm s}^{-1}\,{\rm\AA}^{-1}
\else erg\,cm$^{-2}$\,s$^{-1}$\,\AA$^{-1}$\fi}
\newcommand{  \ergcmsHz  }{\ifmmode{\rm erg\,cm}^{-2}\,{\rm s}^{-1}\,{\rm Hz}^{-1}
                       \else ergs\,cm$^{-2}$\,s$^{-1}$\,Hz$^{-1}$\fi}
\newcommand{\kev}	{\ifmmode {\rm keV} \else keV\fi}

\newcommand{\mic}	{\ifmmode {\rm \mu m} \else $\mu$m\fi}
\newcommand{\vFWHM}	{\ifmmode v_{\mbox{\tiny FWHM}} \else $v_{\mbox{\tiny FWHM}}$\fi}
\newcommand{\vBLR}	{\ifmmode v_{\mbox{\tiny BLR}} \else $v_{\mbox{\tiny BLR}}$\fi}
\newcommand{\sigBLR}	{\ifmmode \sigma_{\mbox{\tiny BLR}} \else $\sigma_{\mbox{\tiny BLR}}$\fi}
\newcommand{\vNLR}	{\ifmmode v_{\mbox{\tiny NLR}} \else $v_{\mbox{\tiny NLR}}$\fi}
\newcommand{\tauBLR}	{\ifmmode \tau_{\mbox{\tiny BLR}} \else $\tau_{\mbox{\tiny BLR}}$\fi}

\newcommand{\Hubble}	{\ifmmode {\rm km\,s}^{-1}\,{\rm Mpc}^{-1} \else km\,s$^{-1}$\,Mpc$^{-1}$\fi}
\newcommand{\NDunit}	{\ifmmode {\rm Mpc}^{-3} \else Mpc$^{-3}$\fi}
\newcommand{\LFunit}	{\ifmmode {\rm Mpc}^{-3}\,{\rm mag}^{-1} \else Mpc$^{-3}$\,mag$^{-1}$\fi}
\newcommand{\MFunit}	{\ifmmode {\rm Mpc}^{-3}\,{\rm dex}^{-1} \else Mpc$^{-3}$\,dex$^{-1}$\fi}

\newcommand{\Msun}{\ifmmode M_{\odot} \else $M_{\odot}$\fi}
\newcommand{\Lsun}{\ifmmode L_{\odot} \else $L_{\odot}$\fi}
\newcommand{\Zsun}{\ifmmode Z_{\odot} \else $Z_{\odot}$\fi}
\newcommand{\mpyr}{\ifmmode \Msun\,{\rm yr}^{-1} \else $\Msun\,{\rm yr}^{-1}$\fi}

\newcommand{\qnote}{\ifmmode q_{0} \else $q_{0}$\fi}
\newcommand{\Hnote}{\ifmmode H_{0} \else $H_{0}$\fi}
\newcommand{\hnote}{\ifmmode h_{0} \else $h_{0}$\fi}
\newcommand{\anote}{\ifmmode a_{0} \else $a_{0}$\fi}
\newcommand{\tnote}{\ifmmode t_{0} \else $t_{0}$\fi}

\newcommand{  \Halpha   }{\ifmmode {\rm H}\alpha \else H$\alpha$\fi}
\newcommand{  \ha   	}{\ifmmode {\rm H}\alpha \else H$\alpha$\fi}
\newcommand{  \Hbeta    }{\ifmmode {\rm H}\beta \else H$\beta$\fi}
\newcommand{  \hb    	}{\ifmmode {\rm H}\beta \else H$\beta$\fi}
\newcommand{  \Hgamma   }{\ifmmode {\rm H}\gamma \else H$\gamma$\fi}
\newcommand{  \Hdelta   }{\ifmmode {\rm H}\delta \else H$\delta$\fi}
\newcommand{  \Lya      }{\ifmmode {\rm Ly}\alpha \else Ly$\alpha$\fi}
\newcommand{  \Lyb      }{\ifmmode {\rm Ly}\beta \else Ly$\beta$\fi}
\newcommand{  \Pa       }{\ifmmode {\rm P}\alpha \else P$\alpha$\fi}
\newcommand{  \Pb       }{\ifmmode {\rm P}\beta \else P$\beta$\fi}
\newcommand{  \Bra      }{\ifmmode {\rm Br}\alpha \else Br$\alpha$\fi}
\newcommand{  \Brg      }{\ifmmode {\rm Br}\gamma \else Br$\gamma$\fi}
\newcommand{  \hii      }{\ifmmode {\rm H}\,\textsc{ii} \else H\,\textsc{ii}\fi}
\newcommand{  \hei      }{\ifmmode {\rm He}\,\textsc{i} \else He\,\textsc{i}\fi}
\newcommand{  \heii     }{\ifmmode {\rm He}\,\textsc{ii} \else He\,\textsc{ii}\fi}
\newcommand{  \HeIIuv   }{\ifmmode {\rm He}\,\textsc{ii}\,\lambda1640 \else He\,\textsc{ii}\,$\lambda1640$\fi}
\newcommand{  \HeIIop   }{\ifmmode {\rm He}\,\textsc{ii}\,\lambda4686 \else He\,\textsc{ii}\,$\lambda4686$\fi}
\newcommand{  \cii      }{\ifmmode {\rm C}\,\textsc{ii}  \else C\,\textsc{ii}\fi}
\newcommand{  \ciii     }{\ifmmode {\rm C}\,\textsc{iii}\right] \else C\,\textsc{iii}]\fi}
\newcommand{  \CIII     }{\ifmmode {\rm C}\,\textsc{iii}\right]\,\lambda1909 \else C\,\textsc{iii}]\,$\lambda1909$\fi}
\newcommand{  \civ      }{\ifmmode {\rm C}\,\textsc{iv}  \else C\,\textsc{iv}\fi}
\newcommand{  \CIV      }{\ifmmode {\rm C}\,\textsc{iv}\,\lambda1549 \else C\,\textsc{iv}\,$\lambda1549$\fi}
\newcommand{\nii}{\ifmmode \left[{\rm N}\,\textsc{ii}\right] \else [N\,{\sc ii}]\fi}

\newcommand{  \niii     }{\ifmmode {\rm N}\,\textsc{iii} \else N\,\textsc{iii}\fi}
\newcommand{  \niv      }{\ifmmode {\rm N}\,\textsc{iv}  \else N\,\textsc{iv}\fi}
\newcommand{  \NIVuv    }{\ifmmode {\rm N}\,\textsc{iv}\,\lambda1486 \else N\,\textsc{iv}\,$\lambda1486$\fi}
\newcommand{  \nv       }{\ifmmode {\rm N}\,\textsc{v}   \else N\,\textsc{v}\fi}
\newcommand{\oi}{\ifmmode \left[{\rm O}\,\textsc{i}\right] \else [O\,{\sc i}]\fi}
\newcommand{\OI}{\ifmmode \left[{\rm O}\,\textsc{i}\right]\,\lambda6300 \else [O\,{\sc i}]$\,\lambda6300$\fi}

\newcommand{\oii}{\ifmmode \left[{\rm O}\,\textsc{ii}\right] \else [O\,{\sc ii}]\fi}
\newcommand{\OII}{\ifmmode \left[{\rm O}\,\textsc{ii}\right]\,\lambda3727 \else [O\,{\sc ii}]\,$\lambda3727$\fi}
\newcommand{\oiii}{\ifmmode \left[{\rm O}\,\textsc{iii}\right] \else [O\,{\sc iii}]\fi}
\newcommand{\OIII}{\ifmmode \left[{\rm O}\,\textsc{iii}\right]\,\lambda5007 \else [O\,{\sc iii}]\,$\lambda5007$\fi}

\newcommand{\NII}{\ifmmode \left[{\rm N}\,\textsc{ii}\right]\,\lambda6583 \else [N\,{\sc ii}]$\,\lambda6583$\fi}

\newcommand{\NeIII}{\ifmmode \left[{\rm Ne}\,\textsc{iii}\right]\,\lambda3968 \else [Ne\,{\sc iii}]$\,\lambda3968$\fi}

\newcommand{\NeV}{\ifmmode \left[{\rm Ne}\,\textsc{v}\right]\,\lambda3426 \else [Ne\,{\sc v}]$\,\lambda3426$\fi}

\newcommand{\HeII}{\ifmmode {\rm He}\,\textsc{ii}\,\lambda4686 \else He\,{\sc ii}$\,\lambda4686$\fi}

\newcommand{\sii}{\ifmmode \left[{\rm S}\,\textsc{ii}\right] \else [S\,{\sc ii}]\fi}

\newcommand{\SII}{\ifmmode \left[{\rm S}\,\textsc{ii}\right]\,\lambda6717,6731 \else [S\,{\sc ii}]$\,\lambda6717,6731$\fi}

\newcommand{  \OIIIuv   }{\ifmmode {\rm O}\,\textsc{iii}\,\lambda1663 \else O\,\textsc{iii}\,$\lambda1663$\fi}
\newcommand{  \oiv      }{\ifmmode {\rm O}\,\textsc{iv}  \else O\,\textsc{iv}\fi}
\newcommand{  \OIVuv    }{\ifmmode {\rm O}\,\textsc{iv}\,\lambda1402  \else O\,\textsc{iv}\,$\lambda1402$\fi}
\newcommand{  \OIVIR    }{\ifmmode {\rm O}\,\textsc{iv}\,25.9\,\mu {\rm m} \else O\,\textsc{iv}\,$25.9\,\mu$m\fi}
\newcommand{  \ovi      }{\ifmmode {\rm O}\,\textsc{vi}   \else O\,\textsc{vi}\fi}
\newcommand{  \Ovi      }{\ifmmode {\rm O}\,\textsc{vi}\,\lambda1035 \else O\,\textsc{vi}\,$\lambda1035$\fi}
\newcommand{  \nei      }{\ifmmode {\rm Ne}\,\textsc{i}   \else Ne\,\textsc{i}\fi}
\newcommand{  \neii     }{\ifmmode {\rm Ne}\,\textsc{ii}  \else Ne\,\textsc{ii}\fi}
\newcommand{  \NeiiIR   }{\ifmmode {\rm Ne}\,\textsc{ii}\,12.8\,\mu {\rm m} \else Ne\,\textsc{ii}\,$12.8\,\mu$m\fi}
\newcommand{  \neiii    }{\ifmmode {\rm Ne}\,\textsc{iii} \else Ne\,\textsc{iii}\fi}
\newcommand{  \neiv     }{\ifmmode {\rm Ne}\,\textsc{iv}  \else Ne\,\textsc{iv}\fi}
\newcommand{  \nev      }{\ifmmode {\rm Ne}\,\textsc{v}   \else Ne\,\textsc{v}\fi}
\newcommand{  \NevIR    }{\ifmmode {\rm Ne}\,\textsc{v}\,24.3\,\mu {\rm m} \else Ne\,\textsc{v}\,$24.3\,\mu$m\fi}
\newcommand{  \nevi     }{\ifmmode {\rm Ne}\,\textsc{vi}  \else Ne\,\textsc{vi}\fi}
\newcommand{  \mgi      }{\ifmmode {\rm Mg}\,\textsc{i}   \else Mg\,\textsc{i}\fi}
\newcommand{  \mgii     }{\ifmmode {\rm Mg}\,\textsc{ii}  \else Mg\,\textsc{ii}\fi}
\newcommand{  \MgII     }{\ifmmode {\rm Mg}\,\textsc{ii}\,\lambda2798 \else Mg\,\textsc{ii}\,$\lambda2798$\fi}
\newcommand{  \siii     }{\ifmmode {\rm S}\,\textsc{iii} \else S\,\textsc{iii}\fi}
\newcommand{  \siv      }{\ifmmode {\rm S}\,\textsc{iv}  \else S\,\textsc{iv}\fi}
\newcommand{  \sili     }{\ifmmode {\rm Si}\,\textsc{i}   \else Si\,\textsc{i}\fi}
\newcommand{  \silii    }{\ifmmode {\rm Si}\,\textsc{ii}  \else Si\,\textsc{ii}\fi}
\newcommand{  \Siliv    }{\ifmmode {\rm Si}\,\textsc{iv}  \else Si\,\textsc{iv}\fi}
\newcommand{  \SilIVuv  }{\ifmmode {\rm Si}\,\textsc{iv}\,\lambda1400  \else Si\,\textsc{iv}\,$\lambda1400$\fi}

\newcommand{  \caii     }{\ifmmode {\rm Ca}\,\textsc{ii}   \else Ca\,\textsc{ii}\fi}
\newcommand{  \feii     }{\ifmmode {\rm Fe}\,\textsc{ii}  \else Fe\,\textsc{ii}\fi}
\newcommand{  \feiii    }{\ifmmode {\rm Fe}\,\textsc{iii} \else Fe\,\textsc{iii}\fi}

\newcommand{ \Lhb   }{\ifmmode L\left(\hb\right) \else $L\left(\hb\right)$\fi}
\newcommand{ \fwhb  }{\ifmmode {\rm FWHM}\left(\hb\right) \else FWHM(\hb)\fi}
\newcommand{ \Lha   }{\ifmmode L\left(\ha\right) \else $L\left(\ha\right)$\fi}
\newcommand{ \fwha  }{\ifmmode {\rm FWHM}\left(\ha\right) \else FWHM(\ha)\fi}
\newcommand{ \Lmg   }{\ifmmode L\left(\mgii\right) \else $L\left(\mgii\right)$\fi}
\newcommand{ \fwmg  }{\ifmmode {\rm FWHM}\left(\mgii\right) \else FWHM(\mgii)\fi}
\newcommand{ \Lciv  }{\ifmmode L\left(\civ\right) \else $L\left(\civ\right)$\fi}
\newcommand{ \fwciv }{\ifmmode {\rm FWHM}\left(\civ\right) \else FWHM(\civ)\fi}
\newcommand{ \fwhm  }{\ifmmode {\rm FWHM} \else FWHM\fi} 
\newcommand{ \voff  }{\ifmmode v_{\rm off} \else $v_{\rm off}$\fi} 

\newcommand{ \mumg  }{\ifmmode \mu\left(\mgii\right) \else $\mu\left(\mgii\right)$\fi}
\newcommand{ \fmg   }{\ifmmode f\left(\mgii\right) \else $f\left(\mgii\right)$\fi}
\newcommand{ \muciv }{\ifmmode \mu\left(\civ\right) \else $\mu\left(\civ\right)$\fi}
\newcommand{ \fciv  }{\ifmmode f\left(\civ\right) \else $f\left(\civ\right)$\fi}
\newcommand{  \auvo     }{\ifmmode \alpha_{\nu,{\rm UVO}} \else $\alpha_{\nu,{\rm UVO}}$\fi}
\newcommand{  \Ledd     }{\ifmmode L_{\rm Edd} \else $L_{\rm Edd}$\fi}
\newcommand{  \lamLlam  }{\ifmmode \lambda L_{\lambda} \else $\lambda L_{\lambda}$\fi}
\newcommand{  \lLl      }{\ifmmode \lambda L_{\lambda} \else $\lambda L_{\lambda}$\fi}
\newcommand{  \nuLnu    }{\ifmmode \nu L_{\nu} \else $\nu L_{\nu}$\fi}
\newcommand{  \nLn      }{\ifmmode \nu L_{\nu} \else $\nu L_{\nu}$\fi}
\newcommand{  \Luv      }{\ifmmode L_{1450} \else $L_{1450}$\fi}
\newcommand{  \Lop      }{\ifmmode L_{5100} \else $L_{5100}$\fi}
\newcommand{  \lLop     }{\ifmmode \log\left(\Lop/\ergs\right) \else $\log\left(\Lop/\ergs\right)$\fi}
\newcommand{  \Lthree   }{\ifmmode L_{3000} \else $L_{3000}$\fi}
\newcommand{  \lLthree  }{\ifmmode \log\left(\Lthree/\ergs\right) \else $\log\left(\Lthree/\ergs\right)$\fi}

\newcommand{\Fthree}{\ifmmode F_{3000} \else $F_{3000}$\fi}
\newcommand{\fuv}{\ifmmode f_{\lambda}\left(1450{\rm \AA}\right) \else $f_{\lambda}\left(1450 {\rm \AA}\right)$\fi}
\newcommand{\fthree}{\ifmmode f_{\lambda}\left(3000{\rm \AA}\right) \else $f_{\lambda}\left(3000{\rm \AA}\right)$\fi}
\newcommand{\fH}{\ifmmode f_{\lambda}\left(1.65\micron\right) \else
$f_{\lambda}\left(1.65\micron\right)$\fi}

\newcommand{\fbol}{\ifmmode f_{\rm bol} \else $f_{\rm bol}$\fi}
\newcommand{\fbolwv}{\ifmmode f_{\rm bol}\left(\lambda\right) \else $f_{\rm bol}\left(\lambda\right)$\fi}
\newcommand{\fbolopt}{\ifmmode f_{\rm bol}\left(5100{\rm \AA}\right) \else $f_{\rm bol}\left(5100{\rm \AA}\right)$\fi}
\newcommand{\fbolthree}{\ifmmode f_{\rm bol}\left(3000{\rm \AA}\right) \else $f_{\rm bol}\left(3000{\rm \AA}\right)$\fi}
\newcommand{\fboluv}{\ifmmode f_{\rm bol}\left(1450{\rm \AA}\right) \else $f_{\rm bol}\left(1450{\rm \AA}\right)$\fi}

\newcommand{  \mbh      }{\ifmmode M_{\rm BH} \else $M_{\rm BH}$\fi}
\newcommand{  \lmbh     }{\ifmmode \log\left(\mbh/\Msun\right) \else $\log\left(\mbh/\Msun\right)$\fi} 
\newcommand{  \lledd    }{\ifmmode L/L_{\rm Edd} \else $L/L_{\rm Edd}$\fi}
\newcommand{  \Lbol     }{\ifmmode L_{\rm bol} \else $L_{\rm bol}$\fi}
\newcommand{  \lbol     }{\ifmmode L_{\rm bol} \else $L_{\rm bol}$\fi}
\newcommand{  \lLbol    }{\ifmmode \log\left(\Lbol/\ergs\right) \else $\log\left(\Lbol/\ergs\right)$\fi} 
\newcommand{  \Lagn     }{\ifmmode L_{\rm AGN} \else $L_{\rm AGN}$\fi}
\newcommand{  \lagn     }{\ifmmode L_{\rm AGN} \else $L_{\rm AGN}$\fi}

\newcommand{  \tgrow     }{\ifmmode t_{\rm growth} \else $t_{\rm growth}$\fi}
\newcommand{  \tUni      }{\ifmmode t_{\rm Universe} \else $t_{\rm Universe}$\fi}

\newcommand{  \Mindot	}{\ifmmode \dot{M}_{\rm infall} \else $\dot{M}_{\rm infall}$\fi}
\newcommand{  \Mbhdot	}{\ifmmode \dot{M}_{\rm BH} \else $\dot{M}_{\rm BH}$\fi}
\newcommand{  \Maddot	}{\ifmmode \dot{M}_{\rm AD} \else $\dot{M}_{\rm AD}$\fi}

\newcommand{  \as	}{\ifmmode a_{\rm *} 		\else $a_{\rm *}$\fi}
\newcommand{  \avec	}{\ifmmode \vec{a}_{\rm *} 	\else $\vec{a}_{\rm *}$\fi}
\newcommand{  \re	}{\ifmmode \eta      	\else $\eta$\fi}

\newcommand{  \mseed    }{\ifmmode M_{\rm seed} \else $M_{\rm seed}$\fi}
\newcommand{  \mbul     }{\ifmmode M_{\rm Bulge} \else $M_{\rm Bulge}$\fi} 
\newcommand{  \mstar    }{\ifmmode M_{*} \else $M_{*}$\fi} 
\newcommand{  \mgal     }{\ifmmode M_{*} \else $M_{*}$\fi} 
\newcommand{  \mhost    }{\ifmmode M_{\rm Host} \else $M_{\rm Host}$\fi}
\newcommand{  \mm       }{\ifmmode M_{*}/M_{\rm BH} \else $M_{*}/M_{\rm BH}$\fi}
\newcommand{  \mmsmall  }{\ifmmode M_{\rm BH}/M_{*} \else $M_{\rm BH}/M_{*}$\fi}
\newcommand{  \mmlarge  }{\ifmmode M_{*}/M_{\rm BH} \else $M_{*}/M_{\rm BH}$\fi}
\newcommand{  \mmwp     }{\ifmmode \left(M_{*}/M_{\rm BH}\right) \else $\left(M_{*}/M_{\rm BH}\right)$\fi}
\newcommand{  \ml       }{\ifmmode M_{*}/L_{*} \else $M_{*}/L_{*}$\fi}
\newcommand{  \mlwp     }{\ifmmode \left(M_{*}/L\right) \else $\left(M_{*}/L\right)$\fi}
\newcommand{  \mlk      }{\ifmmode \left(M_{*}/L_{K}\right) \else $\left(M_{*}/L_{K}\right)$\fi}
\newcommand{  \sigs     }{\ifmmode \sigma_{*} \else $\sigma_{*}$\fi}
\newcommand{  \Reff     }{\ifmmode R_{\rm e} \else $R_{\rm e}$\fi}

\def \nustar {{\em NuSTAR }}
\def \nustarsh {{\em NuSTAR}}
\def \swift {{\em Swift\ }}
\def \swiftxrt {{\em Swift}/XRT\ }
\def \swiftxrtsh {{\em Swift}/XRT}
\def \swiftbat {{\em Swift}/BAT\ }
\def \swiftbatsh {{\em Swift}/BAT}
\def \chandra {{\em Chandra\ }}

\def \xmmsh{{\em XMM-Newton}}
\def \xmm{{\em XMM-Newton\ }}

\def \suzaku{{\em Suzaku\ }}
\def \suzakush{{\em Suzaku}}

\newcommand{\bj}{\ifmmode b_{\rm J} \else $b_{\rm J}$\fi}

\newcommand{\iab}{\ifmmode i_{\rm AB} \else $i_{\rm AB}$\fi}

\newcommand{\jab}{\ifmmode J_{\rm AB} \else $J_{\rm AB}$\fi}
\newcommand{\hab}{\ifmmode H_{\rm AB} \else $H_{\rm AB}$\fi}
\newcommand{\kab}{\ifmmode K_{\rm AB} \else $K_{\rm AB}$\fi}

\newcommand{\jveg}{\ifmmode J_{\rm Vega} \else $J_{\rm Vega}$\fi}
\newcommand{\hveg}{\ifmmode H_{\rm Vega} \else $H_{\rm Vega}$\fi}
\newcommand{\kveg}{\ifmmode K_{\rm Vega} \else $K_{\rm Vega}$\fi}

\def\arcmin{\hbox{$^\prime$}}
\def\arcsec{\hbox{$^{\prime\prime}$}}

\newcommand{  \Chisq    }{\ifmmode \chi^{2} \else $\chi^{2}$}
\newcommand{  \nelec    }{\ifmmode n_{e} \else $n_{e}$\fi}     % electron density
\newcommand{  \nh       }{\ifmmode n_{H} \else $n_{H}$\fi}     % hydrogen density
\newcommand{  \Ncol     }{\ifmmode N_{col} \else $N_{col}$\fi} % column density
\newcommand{  \NH       }{\ifmmode N_{H} \else $N_{\rm H}$\fi}     % column density

\def\deg{\hbox{$^\circ$}}

\def\arcmin{\hbox{$^\prime$}}
\def\arcsec{\hbox{$^{\prime\prime}$}}
\def\ion#1#2{#1$\;${\small\rm\@Roman{#2}}\relax}

\newcommand{\OIIIa}{\ifmmode \left[{\rm O}\,\textsc{iii}\right]\,\lambda4959 \else [O\,{\sc iii}]\,$\lambda4959$\fi}
\newcommand{\NIIa}{\ifmmode \left[{\rm N}\,\textsc{ii}\right]\,\lambda6548 \else [N\,{\sc ii}]\,$\lambda6548$\fi}
\newcommand{\SIIa}{\ifmmode \left[{\rm S}\,\textsc{ii}\right]\,\lambda6716 \else [S\,{\sc ii}]\,$\lambda6716$\fi}
\newcommand{\SIIb}{\ifmmode \left[{\rm S}\,\textsc{ii}\right]\,\lambda6732 \else [S\,{\sc ii}]\,$\lambda6731$\fi}
\newcommand{\NeVa}{\ifmmode \left[{\rm Ne}\,\textsc{v}\right]\,\lambda3346 \else [Ne\,{\sc v}]\,$\lambda3346$\fi}
\newcommand{\NeVb}{\ifmmode \left[{\rm Ne}\,\textsc{v}\right]\,\lambda3426 \else [Ne\,{\sc v}]\,$\lambda3426$\fi}
\newcommand{\NeIIIa}{\ifmmode \left[{\rm Ne}\,\textsc{iii}\right]\,\lambda3869 \else [Ne\,{\sc iii}]\,$\lambda3869$\fi}
\newcommand{\NeIIIb}{\ifmmode \left[{\rm Ne}\,\textsc{iii}\right]\,\lambda3968 \else [Ne\,{\sc iii}]\,$\lambda3968$\fi}
\newcommand{\Mgb}{\ifmmode \left{\rm Mg}\,\textsc{i}\right\,\lambda5175 \else Mg\,{\sc i}\,$\lambda5175$\fi}

\newcommand{\mgb}{\ifmmode \left{\rm Mg}\,\textsc{i}\right \else Mg\,{\sc i}\fi}

\newcommand{\Cahk}{\ifmmode \left[{\rm Ca H+K}\,\textsc{ii}\right]\,\lambda3935,4214 \else [Ca H+K]$\,\lambda3935,4214$\fi}

\def\kmpssh{\hbox{$\km\s^{-1}$}}
\def\ergpssh{\hbox{$\erg\s^{-1}$}}

\def\arcmin{{\mbox{$^{\prime}$}}}
\def\degree{{\mbox{$^{\circ}$}}}
\def\arcsec{{\mbox{$^{\prime \prime}$}}}

\def\erg{{\rm\thinspace erg}}

\def\keV{{\rm\thinspace keV}}
\def\km{{\rm\thinspace km}}

\def\Lsun{\hbox{$\rm\thinspace L_{\odot}$}}

\def\Msun{\hbox{$\rm\thinspace M_{\odot}$}}
\def\pc{{\rm\thinspace pc}}

\def\s{{\rm\thinspace s}}

\def \mytorus {{\tt MYtorus\ }}
\newcommand{\angstrom}{\mbox{\normalfont\AA}}

\def \ngc {NGC\,6921}
\def \mcg {MCG\,$+$04-48-002}
\newcommand {\Lsoft} {$L_{\mathrm{2-10\ keV}}$}

\newcommand {\nhunit} {cm$^{-2}$}
\newcommand {\ergpersec} {erg~s$^{-1}$}

\usepackage[T1]{fontenc}
\usepackage{babel}

\begin{document}

\title{\nustar\ Resolves the First Dual AGN above 10 keV in SWIFT\,J2028.5$+$2543} 

\author{Michael J. Koss\altaffilmark{1,17}, Ana Glidden\altaffilmark{2}, Mislav~Balokovi{\' c}\altaffilmark{3}, Daniel~Stern\altaffilmark{4}, Isabella Lamperti\altaffilmark{1}, Roberto Assef\altaffilmark{5},  Franz Bauer\altaffilmark{6,7}, David Ballantyne\altaffilmark{8}, Steven E. Boggs\altaffilmark{9}, William W. Craig\altaffilmark{9,10}, Duncan Farrah\altaffilmark{11}, Felix F{\"u}rst\altaffilmark{3}, Poshak Gandhi\altaffilmark{12}, Neil Gehrels\altaffilmark{13}, Charles J. Hailey\altaffilmark{14}, Fiona A. Harrison\altaffilmark{3}, Craig Markwardt\altaffilmark{13}, Alberto Masini\altaffilmark{15}, Claudio Ricci\altaffilmark{3}, Ezequiel Treister\altaffilmark{6,16}, Dominic J. Walton\altaffilmark{4,3}, and William W. Zhang\altaffilmark{13}}

\altaffiltext{1}{Institute for Astronomy, Department of Physics, ETH Zurich, Wolfgang-Pauli-Strasse 27, CH-8093 Zurich, Switzerland; mkoss@phys.ethz.ch}
\altaffiltext{2}{Massachusetts Institute of Technology, 77 Massachusetts Ave, Cambridge, MA 02139, USA}
\altaffiltext{3}{Cahill Center for Astronomy and Astrophysics, California Institute of Technology, Pasadena, CA 91125, USA}
\altaffiltext{4}{Jet Propulsion Laboratory, California Institute of Technology, 4800 Oak Grove Drive, Mail Stop 169-221, Pasadena,CA 91109, USA}
\altaffiltext{5}{N\'ucleo de Astronom\'ia de la Facultad de Ingenier\'ia, Universidad Diego Portales, Av. Ej\'ercito 441, Santiago, Chile}
\altaffiltext{6}{Instituto de Astrof\'{\i}sica, Facultad de F\'{\i}sica, Pontificia Universidad Cat\'olica de Chile, Casilla 306, Santiago 22, Chile}
\altaffiltext{7}{Space Science Institute, 4750 Walnut Street, Suite 205, Boulder, CO 80301, USA}
\altaffiltext{8}{Center for Relativistic Astrophysics, School of Physics, Georgia Institute of Technology, Atlanta, GA 30332, USA}
\altaffiltext{9}{Space Sciences Laboratory, 7 Gauss Way, University of California, Berkeley, CA 94720-7450, USA}
\altaffiltext{10}{Lawrence Livermore National Laboratory, Livermore, CA 94550, USA}
\altaffiltext{11}{Department of Physics, Virginia Tech, Blacksburg, VA, 24061, USA}
\altaffiltext{12}{School of Physics and Astronomy, University of Southampton, Highfield, Southampton SO17 1BJ, UK}
\altaffiltext{13}{NASA Goddard Space Flight Center, Greenbelt, MD 20771, USA}
\altaffiltext{14}{Columbia Astrophysics Laboratory, Columbia University, New York, NY 10027, USA}
\altaffiltext{15}{INAF--Osservatorio Astronomico di Bologna, via Ranzani 1, 40127 Bologna, Italy}
\altaffiltext{16}{Departamento de Astronom\'{\i}a, Universidad de Concepci\'{o}n, Concepci\'{o}n, Chile}
\altaffiltext{17}{SNSF Ambizione Fellow}
\email{mkoss@phys.ethz.ch}

%%%%%ABSTRACT%%%%%%%

\begin{abstract}
We have discovered heavy obscuration in the dual active galactic nucleus (AGN) in the \swiftbat source SWIFT\,J2028.5$+$2543 using \nustarsh.   While an early \xmm study suggested the emission was mainly from \ngc, the superior spatial resolution of \nustar above 10 keV resolves the \swiftbat emission into two sources associated with the nearby galaxies \mcg\ and \ngc\ ($z=0.014$) with a projected separation of 25.3\,kpc (91\arcsec).  \nustarsh's sensitivity above 10\,keV finds both are heavily obscured to Compton-thick ($N_{\rm H}\approx1-2\times10^{24}$\,\nhunit) and contribute equally to the BAT detection ($L_{\rm 10-50 \ keV}^{\rm int}\approx6\times10^{42}$\,\ergpssh).  The observed luminosity of both sources is severely diminished in the 2-10 keV band ($L_{\rm \ 2-10 \ keV}^{\rm obs}<0.1\times L_{\rm \ 2-10 \ keV}^{\rm int}$), illustrating the importance of $>10$ keV surveys like those with \nustar and \swiftbatsh. Compared to archival X-ray data, \mcg\ shows significant variability ($>$3) between observations. Despite being bright,  X-ray detected AGN, both are difficult to detect using optical emission line diagnostics because \mcg\ is identified as a starburst/composite because of the high rates of star formation from a luminous infrared galaxy while \ngc\ is only classified as a LINER using line detection limits.  SWIFT\,J2028.5$+$2543 is the first dual AGN resolved above 10 keV and is the 2nd most heavily obscured dual AGN discovered to date in the X-rays other than NGC 6240. 
 \end{abstract}


%%%%%INTRODUCTION%%%%%%%
\section{Introduction}
% old: To further understand the physics of AGN, {\it NuSTAR} (\citealt{Harrison}) is undertaking a snapshot survey ($\sim 20$~ks) of bright, local AGN selected from the \emph{Swift}-Burst Alert Telescope (BAT; REF) survey (\citealt{Cusumano}).  Targeted snapshot follow-up of BAT sources with {\it NuSTAR} provide sensitive broad-band spectra of sources, which previously only had moderate significance ($< 10 \sigma$) BAT detections.  

% old: As the first X-ray satellite in orbit to focus X-rays at energies significantly above 10~keV, {\it NuSTAR} has approximately an order of magnitude improved spatial resolution and two orders of magnitude improved sensitivity compared to previous hard X-ray telescopes such as BAT, albeit with a more modest, $\sim 13\arcmin \times 13\arcmin$ field of view.  {\it NuSTAR} has revealed a large population of heavily-obscured AGN, in line with the 80\% obscuration rate of BAT AGN (\citealt{Burlon}).  For the first time, we will have a good sample to study the obscured AGN that make up the CXB.

%Inspired by the early observational work of \citet{Sanders:1984:182}, theorists often invoke galaxy mergers to fuel supermassive black holes by efficiently transporting gas into the nucleus \citep[e.g.,][]{Barnes:1991:40}.  

Over the last decade, dual active galactic nuclei (AGN) have been found serendipitously \citep[e.g.,][]{Komossa:2003:L15,Koss:2011:L42,Comerford:2011:L19a} and also through large systematic surveys using optical spectroscopy \citep[e.g.,][]{Liu:2011:101,Comerford:2013:64}, X-ray emission \citep[][]{Koss:2012:L22,Liu:2013:110,Comerford:2015:219}, or radio observations \citep[][]{Fu:2015:72, MullerSanchez:2015:103}.    This work has suggested that close ($<$30 kpc) major galaxy mergers are efficient at triggering AGN. Theorists have also suggested that AGN obscuration can rise to Compton-thick levels in the merging process \citep[\NH$>$$10^{24}$\,\nhunit,][]{Hopkins:2005:L71}.  However, only one dual has been found where both AGN are Compton-thick: NGC 6240 \citep[][]{Komossa:2003:L15}.  \\

%, which was discovered serendipitously  
%This is a consequence of the limited sensitivity of \chandra above 2 keV, which makes estimates of column density difficult to measure at high levels of obscuration.\\

The \swift Burst Alert Telescope \citep[BAT,][]{Barthelmy:2005:143}  has proven important in nearby obscured AGN studies, because it is sensitive to the 14--195\,\keV\ band, and thus to X-rays which can penetrate through even Compton-thick columns of obscuring material (\NH$>10^{24}$\,\nhunit).  Studies of BAT-detected AGN have suggested that this sensitivity is linked to the high fraction of mergers and dual AGN \citep{Koss:2010:L125,Koss:2012:L22}. Unfortunately, the limited angular resolution (FWHM$\approx20\arcmin$) and large positional uncertainty \citep[$\approx3\arcmin$,][]{Baumgartner:2013:19} make \swiftbat ill suited for dual AGN studies because of source confusion.  With the new high-energy focusing optics on the {\it Nuclear Spectroscopic Telescope Array} \citep[{\it NuSTAR};][]{Harrison:2013:103}, the 3-79 keV energy range can be studied at sensitivities and angular resolutions 10--100 times better than \swiftbatsh.   Additionally, the $>$10 keV sensitivity of \nustar has found intrinsic (unabsorbed) X-ray luminosities can be $\approx$10-70 times higher for heavily obscured sources than pre-NuSTAR constraints from \chandra or \xmm \citep[][]{Lansbury:2015:115}.  \\

%Unfortunately, the limited angular resolution (FWHM$\approx20\arcmin$) and large positional uncertainty \citep[$\approx3\arcmin$,][]{Baumgartner:2013:19} \swiftbat is ill suited for dual AGN studies because of source confusion. 
%which are 5--20$\times$ larger than studies of nearby AGN selected at other wavelengths

 \ngc\ and \mcg\ were first found to host possible X-ray counterparts to SWIFT\,J2028.5$+$2543 based on \swiftxrt and \xmm spectra that suggested that NGC 6921 was the primary BAT source \citep{Winter:2008:686} and was nearly Compton-thick ($N_{\rm H}\approx1\times10^{24}$\,\nhunit) while \mcg\ had a complex spectra with no evidence of obscuration.  It was later classified as a dual AGN  \citep[][]{Koss:2012:L22} based on the luminous hard X-ray emission in both AGN (\Lsoft$>10^{42}$\,\ergpssh) and the small redshift ($<500$\,\kmpssh) and physical (25.2 kpc) separation between the host galaxies. A recent compilation of BAT-detected AGN found NGC 6921 was likely Compton-thick \citep{Ricci:2015:L13}.  Here, we use \nustar to resolve the $>$10 keV emission to find a heavily obscured dual AGN pair in the \swiftbat source SWIFT\,J2028.5$+$2543 with both sources contributing equally.  Throughout this Letter, we adopt $\Omega_m=0.3$, $\Omega_\Lambda=0.7$, and $H_0=70$\,km\,s$^{-1}$\,Mpc$^{-1}$.
  
  %As part of the \nustar  \swiftbat snapshot program (Balokovic et al., in prep), we discovered a heavily obscured dual AGN in the BAT source SWIFT\,J2028.5$+$2543 with a 91\arcsec\ (25.3 kpc) separation at a distance of 53\,Mpc.      

%

\section{Observations and Data Reduction}
We describe here the optical imaging and spectroscopy (Section \ref{sec:optical}) and X-ray observations (Section \ref{sec:xray_obs}). Errors are quoted at the 90\% confidence level for the parameter of interest unless otherwise specified.  

%At the redshift of the pair, 1\arcsec\ corresponds to 0.29\,kpc. 

\subsection{Optical Imaging and Spectroscopy} \label{sec:optical}
Optical imaging was obtained in an earlier survey of 185 BAT AGN \citep[{\em ugriz} from][]{Koss:2011:57} using the Kitt Peak 2.1m telescope.     For optical spectroscopy, we used the Double Spectrograph (DBSP) on the Hale 200-inch telescope at Palomar Observatory.   On UT 2013 August 13, we observed \mcg\ for 500~s with a 1\farcs5\ slit and \ngc\ for 300~s with a 0\farcs5 slit both at the parallactic angle (-68\deg). We also observed a nearby galaxy, 2MASX\,J20283767+2543183, for 600~s with a 1\farcs5 slit (Fig.~\ref{fig:opticalimage}) on UT 2015 July 22 at the parallactic angle (58\deg).    We processed the data with flux calibration from observations of BD~+17 3248, BD~+33 2642, and Feige~110.  Milky Way Galactic reddening has been taken into account according to \citet{Schlafly:2011:103}. 

%We follow the procedure of \citet{Koss:2011:57} using SExtractor to measure photometry \citep{Bertin:1996:393}.  
%near-infrared (NIR; 2MASS-{\em JHK}), mid-IR ({\em WISE}, 3.4-22 $\micron$), as well as \swift ultraviolet (UV) imaging We follow the photometry procedure of \citet{Koss:2011:57} using Kitt Peak or 2MASS data to measure the optical and NIR photometry.  We use the \citet{Assef:2010:970} 0.03-30 $\micron$ algorithm to model the strength of the AGN emission in the mid-IR using empirical AGN and galaxy templates.
%Nearby foreground stars and galaxies were identified using segmentation maps produced by .
%obtained during the same observing run with the same instrument configurations The 0\farcs5 and 1\farcs5 slit correspond to physical scales of $\approx$150\,pc and $\approx$450\,pc,  respectively.

We use the penalised PiXel Fitting software \citep[{\tt pPXF},][]{Cappellari:2004:138} to measure stellar kinematics and the central stellar velocity dispersion ($\sigma_\star$) with the Indo-U.S. CaT, and MILES empirical stellar library \citep[$3465-9468\,\angstrom$][]{Vazdekis:2012:157}.   We fit the residual spectra for emission lines after subtracting the stellar templates with the {\tt PYSPECKIT} software following \citet[][]{Berney:2015:3622} and correct the narrow line ratios (\ha/\hb) assuming an intrinsic ratio of R=3.1 and the \citet{Cardelli:1989:245} reddening curve.

%     To correct for intrinsic galactic extinction, we measure .  In the case of a \hb\ non-detection, we assume the 3$\sigma$ upper limits for the extinction correction.
%e fit three separate regions independently corresponding to the blue and red spectra at wavelengths $3900-5200\,\AA$, $5580-6900\,\AA$, and $8460-8700\,\AA$ corresponding to the Calcium triplet absorption lines. We follow the emission line masking procedure used in \citet{Oh:2011:13}. Narrow-line widths were fixed to the \OIII\ line.

%and parameterized the line-of-sight velocity distribution (LOSVD) by means of a simple Gaussian , which uses a Levenberg-Marquardt algorithm for fitting

%at resolutions of $2.5\,\angstrom$ FWHM

\subsection{X-ray Observations} \label{sec:xray_obs}
A summary of the X-ray observations is in Table~\ref{tab:xray_data}.    \nustar observed SWIFT\,J2028.5$+$2543 on 2013 May~18. The data was processed using the {\tt NuSTARDAS} software version 1.4.1 and CALDB version 20150702. The exposure time totaled 19.5\,ks.  Rather than a single bright source, two point sources are seen in the \nustar images.  For spectral extraction, we used circular regions 40\arcsec\ in radius centered on the point-source peaks.   A background spectrum was extracted from a polygonal region surrounding both sources.  The  counts totaled 780 in \mcg\ and 624 in \ngc.  We required at least 20 counts per bin for fitting.

%The extraction region size and spectral binning were optimized for short \nustar observations of obscured AGN, which effectively results in at least 20\,counts per bin after background subtraction, and maximizes signal-to-noise ratio above 10\,keV and spectral resolution in the Fe-complex region (5--8\,keV), as described in M. Balokovi{\' c} et~al., in preparation. background subtracted 

The \nustar observation was coordinated with a \swiftxrt exposure of 6.6\,ks on the same day.  \swiftxrt also observed the system three times in the past. \swiftxrt data was processed using the ASI Science Data Center tools. We used a 71\arcsec\ circular extraction region, and background extraction regions with inner and outer radii of 142\arcsec\ and 236\arcsec, respectively, with a minimum of three counts per bin for fitting.

%, so we co-added all available \swiftxrt spectra resulting in total exposure of 16.6\,ks  \mcg\ significantly varied in count rate between the observation in 2013 and earlier three observations in 2005 and 2006 (when it was higher by a factor of $\simeq$2-3 at the 3-5$\sigma$ level), so we did not combine the 2013 data with the older data because of possible spectral variability. For \ngc\ there is no variation in the source count rate among the four publicly available observations.

The system was previously observed by \xmm and \suzaku on 2006 April~4 and 2007 April~18, respectively. We processed the data using standard procedures\footnote{See \url{http://heasarc.gsfc.nasa.gov/docs/xmm/abc/} and \url{http://www.astro.isas.jaxa.jp/suzaku/process/}.}, employing SAS (version 7.0) for the \xmm EPIC data and the HEAsoft script {\tt aepipeline} for the \suzaku XIS data.

% If needed: XMM extraction region radii were 32 and 19 arcsec. Suzaku extraction regions were 60 and 30 arcsec. Larger is for \mcg in each case, since it's brighter in both. ASDC\footnote{, \url{http://www.asdc.asi.it/mmia/}}

%%%%%%%%%%%%%%%%%%%%%%%%%%%%%%%%%%%%%%%%%%%%%%%%%%%%%%%%%%%%%%%%%%%%%%%%%%%%%%%%%%%%%%%%%%%%%%%%%%%
\begin{deluxetable*}{ccccc} %%%%%%%%%%%%%%%%%%%%%%%%%%%%%%%%%%%%%%%%%%%%%%%%%%%%%%%%%%%%%%%%%%%%%%%
\tabletypesize{\footnotesize}
\tablewidth{0cm}
\renewcommand{\arraystretch}{5}
\tablecaption{ Summary of X-ray Observations \label{tab:xray_data}}

\tablehead{
  \colhead{\multirow{2}{*}{Observatory}} &
  \colhead{\multirow{2}{*}{Observation ID}} &
  \colhead{\multirow{2}{*}{Date}} &
  \colhead{Exp.} &
  \colhead{Source Count Rate\,\tablenotemark{a} (\,s$^{-1}$\,)} \\
  \colhead{} &
  \colhead{} &
  \colhead{} &
  \colhead{(\,ks\,)} &
  \colhead{\mcg\ /\ \ngc}
}

% NOTE: The last column (now commented out) are 2-10 keV fluxes and upper limit estimates in 10^-13 erg/s/cm2 calculated from best-fit models or estimated from count rates. It can be included if the numbers seem useful for discussion -- for now that doesn't seem all that relevant and may require some explaining regarding calculations.
%\startdata %%%%%%%%%%%%%%%%%%%%%%%%%%%%%%%%%%%%%%%%%%%%%%%%%%%%%%%%%%%%%%%%%%%%%%%%%%%%%%%%%%%%%%
\swift\,(XRT) & 00035276001 & 2005 Dec 16 & 4.5 & 0.015\ /\ 0.002 \\ % $20\pm10$ / $<6$
\swift\,(XRT) & 00035276002 & 2006 Mar 23 & 4.6 & 0.013\ /\ 0.004 \\ % $20\pm10$ / $<6$
%\swift\,(XRT) & 00035276003 & 2006 Mar 28 & 0.9 & 0.010\ /\ 0.004 \\ % $20\pm10$ / $<8$
\swift\,(XRT) & 00030722001 & 2006 Jun 3 & 6.9 & 0.011\ /\ 0.003 \\ % $20\pm10$ / $<8$
\swift\,(XRT) & 00080266001 & 2013 May 18 & 6.6 & 0.005\ /\ 0.003 \\ % $7\pm2$ / $<7$
%\swift\,(XRT) & co-added    & 2005--2013 & 16.6 & \nodata\ /\ 0.003 \\ % \nodata / $4\pm1$
\xmm (EPIC) & 0312192301 & 2006 Apr 23 & 8.8 & 0.150\ /\ 0.035 \\ % $26_{-14}^{+1}$ / $5_{-1}^{+4}$ % OK but switched in Winter+2008
\suzaku (XIS) & 702081010 & 2007 Apr 18 & 41.3 & 0.026\ /\ 0.008 \\ % $22_{-13}^{+1}$ / $6_{-5}^{+1}$ % OK with Winter+2009
\nustar   & 60061300002 & 2013 May 18 & 19.5 & 0.040\ /\ 0.032 \\ % $7\pm2$ / $4\pm1$
\swift\,(BAT) & 104 month & 2005-2013 & 10894 & 0.002 
\enddata %%%%%%%%%%%%%%%%%%%%%%%%%%%%%%%%%%%%%%%%%%%%%%%%%%%%%%%%%%%%%%%%%%%%%%%%%%%%%%%%%%%%%%%%

\tablenotetext{a}{Background-subtracted instrument count rate in: 0.3--10\,keV for \swift (XRT) and \xmm (EPIC), 0.1--12\,keV for \suzaku (average between XIS0, XIS1 and XIS3), 3--79\,keV for \nustar (FPMA), and 14--195\,keV for \swift (BAT). The BAT count rate is in Crab units. }
\end{deluxetable*} %%%%%%%%%%%%%%%%%%%%%%%%%%%%%%%%%%%%%%%%%%%%%%%%%%%%%%%%%%%%%%%%%%%%%%%%%%%%%%%
%%%%%%%%%%%%%%%%%%%%%%%%%%%%%%%%%%%%%%%%%%%%%%%%%%%%%%%%%%%%%%%%%%%%%%%%%%%%%%%%%%%%%%%%%%%%%%%%%%

\section{Results}
We first describe results from optical imaging and spectroscopy (Section \ref{sec:optical_fits}), then discuss X-ray variability and spectral modeling (Section \ref{sec:xray_variability} and \ref{sec:xray_fits}).  We  follow with a discussion of the intrinsic AGN luminosity (Section \ref{sec:intrin_lum}).  

\subsection{Optical Imaging and Spectroscopy} \label{sec:optical_fits}
\begin{figure*}
  \begin{center}
    \leavevmode
      \includegraphics[height=9cm]{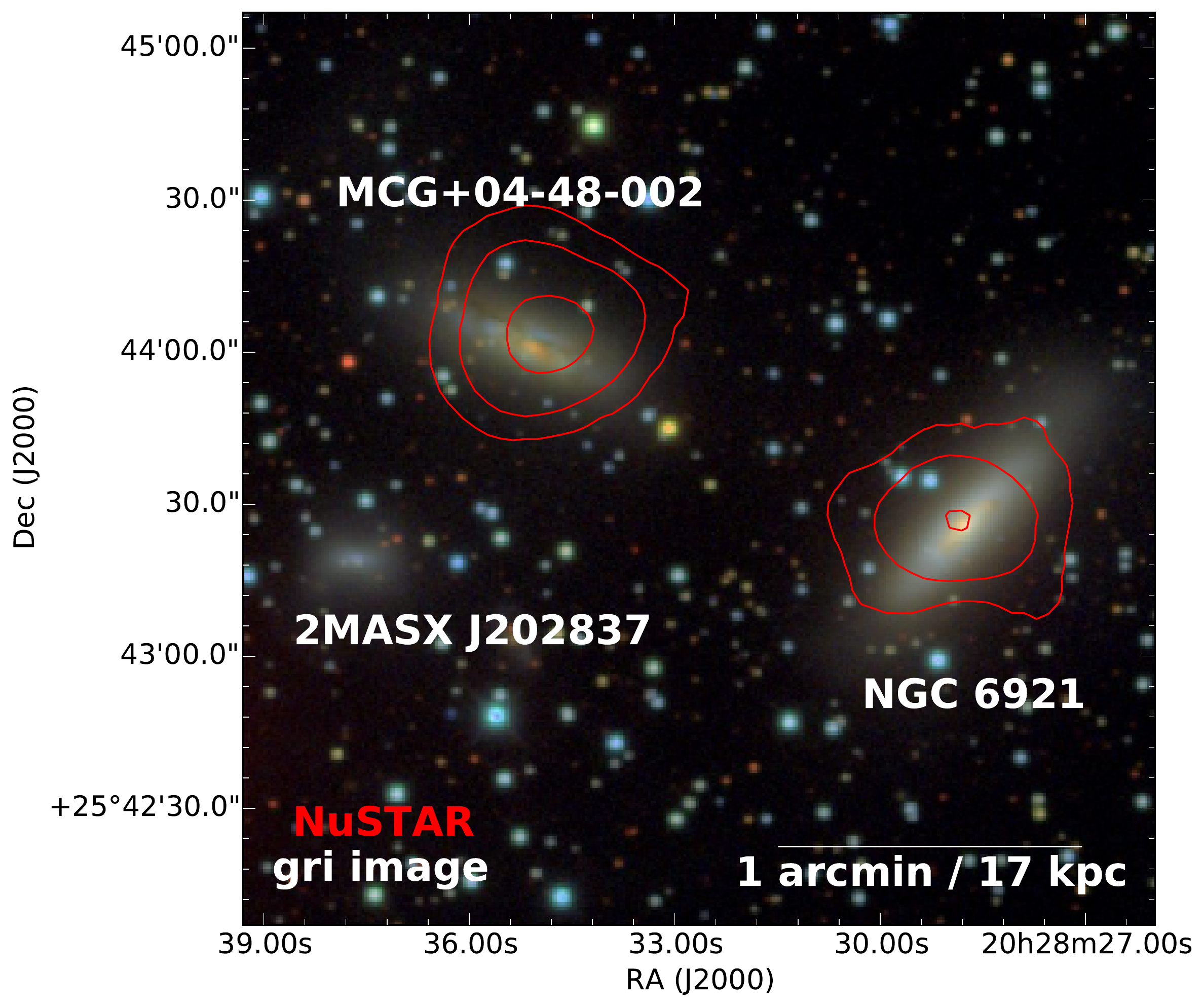}
      \includegraphics[width=7cm]{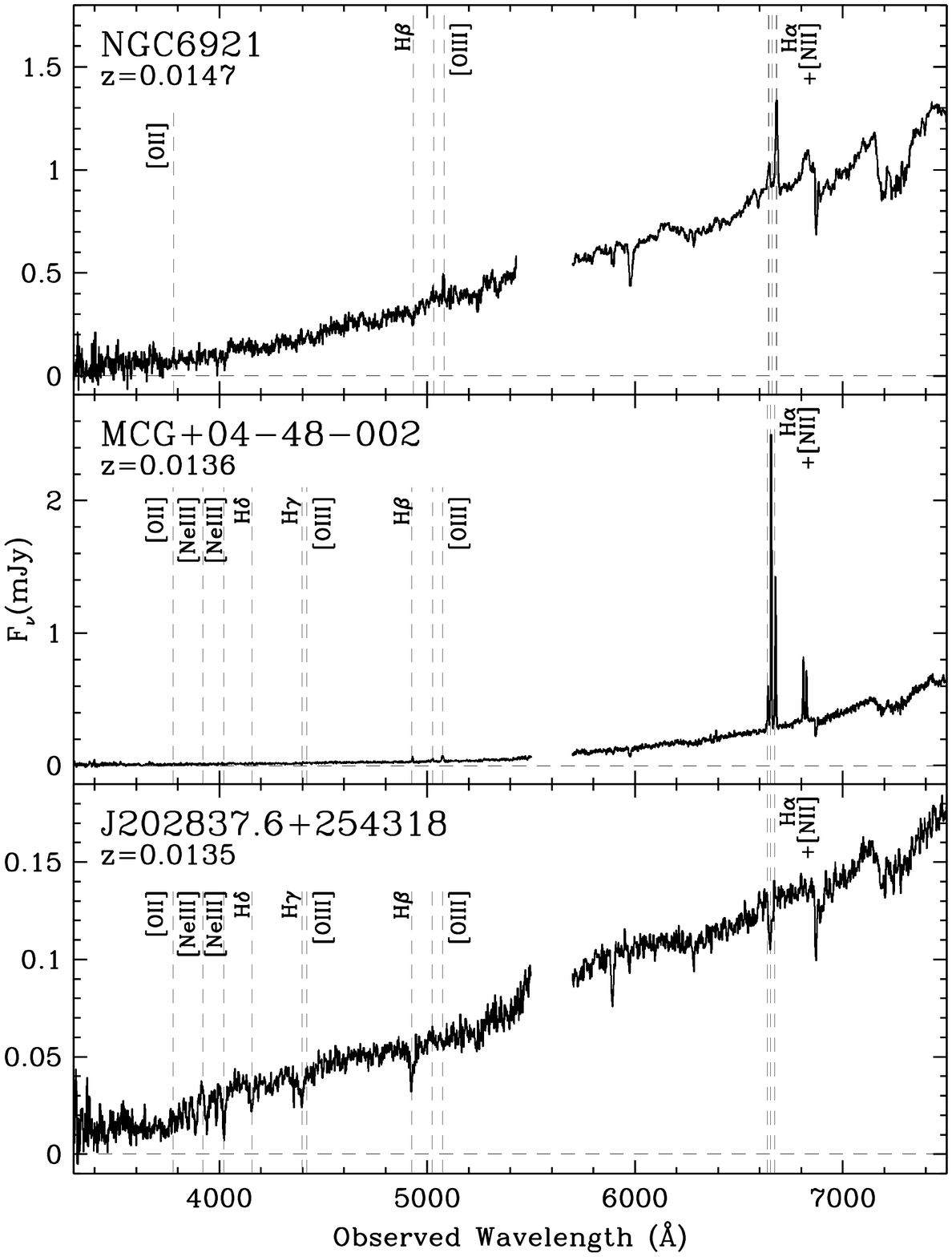}
       \caption[Optical Image]{ {\it Left:} Color Kitt Peak 2.1m optical $gri$ image displayed  with an arcsinh scale with \nustar X-ray (3-79 keV) contours overlaid in red.  2MASX J202837.6+254318, the faintest galaxy in the group, is not detected by \nustarsh.  {\it Right:} Optical spectra of  NGC 6921, MCG +04-48-002, and 2MASX J202837.6+254318 from Palomar.  The three galaxies are found at similar redshifts, suggesting a galaxy group.  Since the system is near the Galactic plane ($\delta_{\rm Gal}=-7^{\circ}$), large foreground Galactic extinction ($\approx$1 mag) and contamination by foreground stars make detection of extended merger features such as tidal tails difficult. }
\label{fig:opticalimage}
  \end{center}
\end{figure*}

A tricolor optical image ($gri$) with \nustar emission overlaid is presented in Fig.~\ref{fig:opticalimage}.  The \OIII emission line is measured at ($z=0.0136$) in \mcg\ and at  ($z=0.0147$) in \ngc.   We also measure the Na I $\lambda\lambda$ 5890, 5896 (Na D) absorption lines from stars and cold gas since narrow emission lines in AGN often have blueshifts compared to their hosts \citep{Bertram:2007:571}.   We measure a restframe velocity of 4212$\pm$15\,km\,s$^{-1}$ ($z=0.0139$) for \mcg\ and 4356$\pm$15\,km\,s$^{-1}$ ($z=0.0141$) for \ngc\  for a $\approx$140\,km\,s$^{-1}$ offset.

There is a 91\arcsec\ separation between \mcg\ and \ngc\ which corresponds to 25.3 kpc at the  \oiii\ line redshift in \mcg\ ($z=0.0136$).  This is slightly larger than the 25.2 kpc separation in \citep[][]{Koss:2012:L22} because of the new DBSP spectra. Imaging shows three additional nearby extended galaxies (major axis$>$20\arcsec) within 360\arcsec\ of \mcg\ (100\,kpc), 2MASX J20283767+2543183 (15.0\,kpc South), 2MASX J20285039+2545324 (62.8\,kpc East), and 2MASX J20283695+2540123 (63.3\,kpc South).  We confirm that 2MASX J20283767+2543183 is an inactive elliptical galaxy at the same redshift ($z$=0.0135) based on the \Hbeta\ absorption. 

\begin{figure*} 
\centering
\includegraphics[width=7cm]{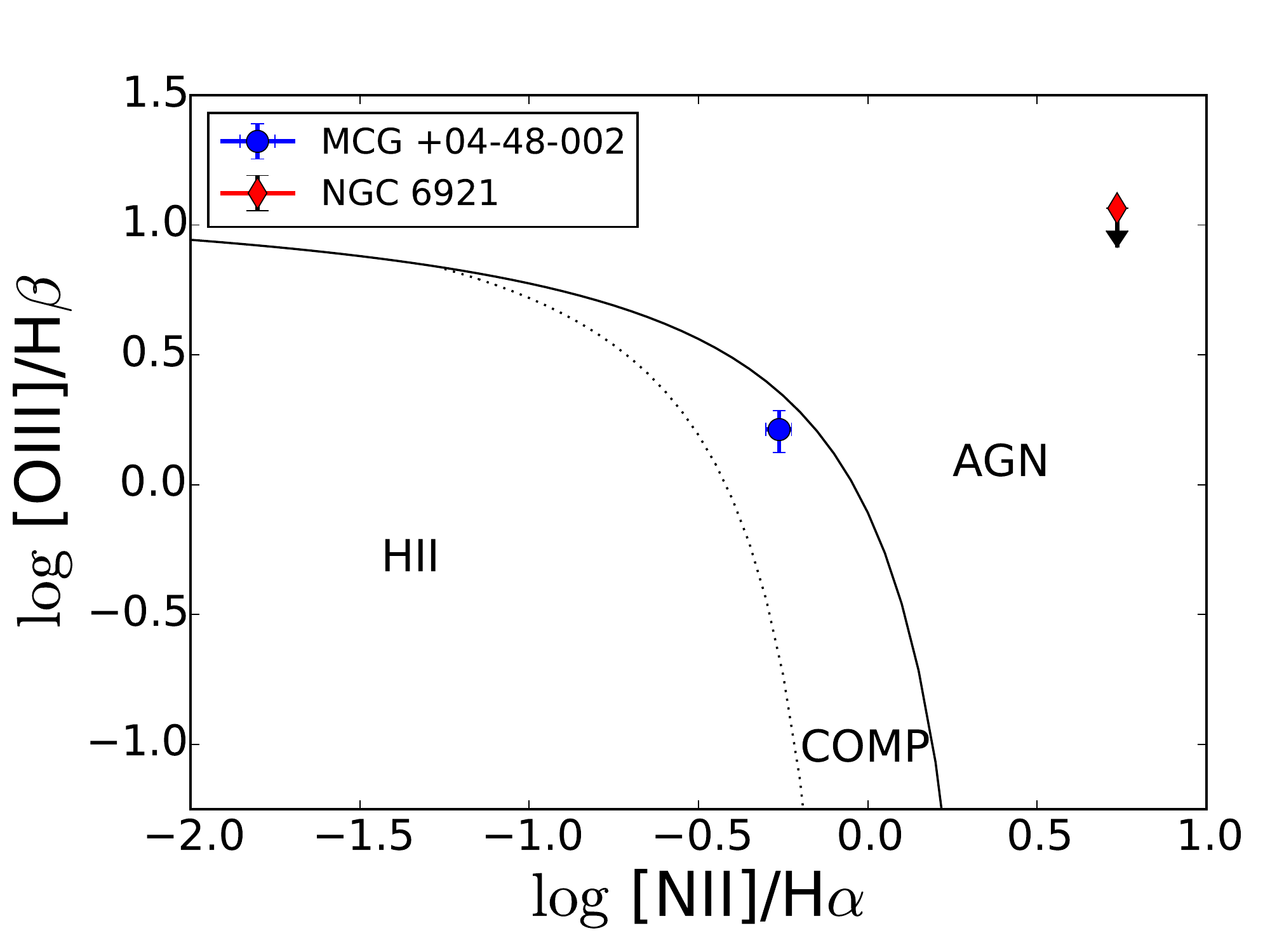}
\includegraphics[width=7cm]{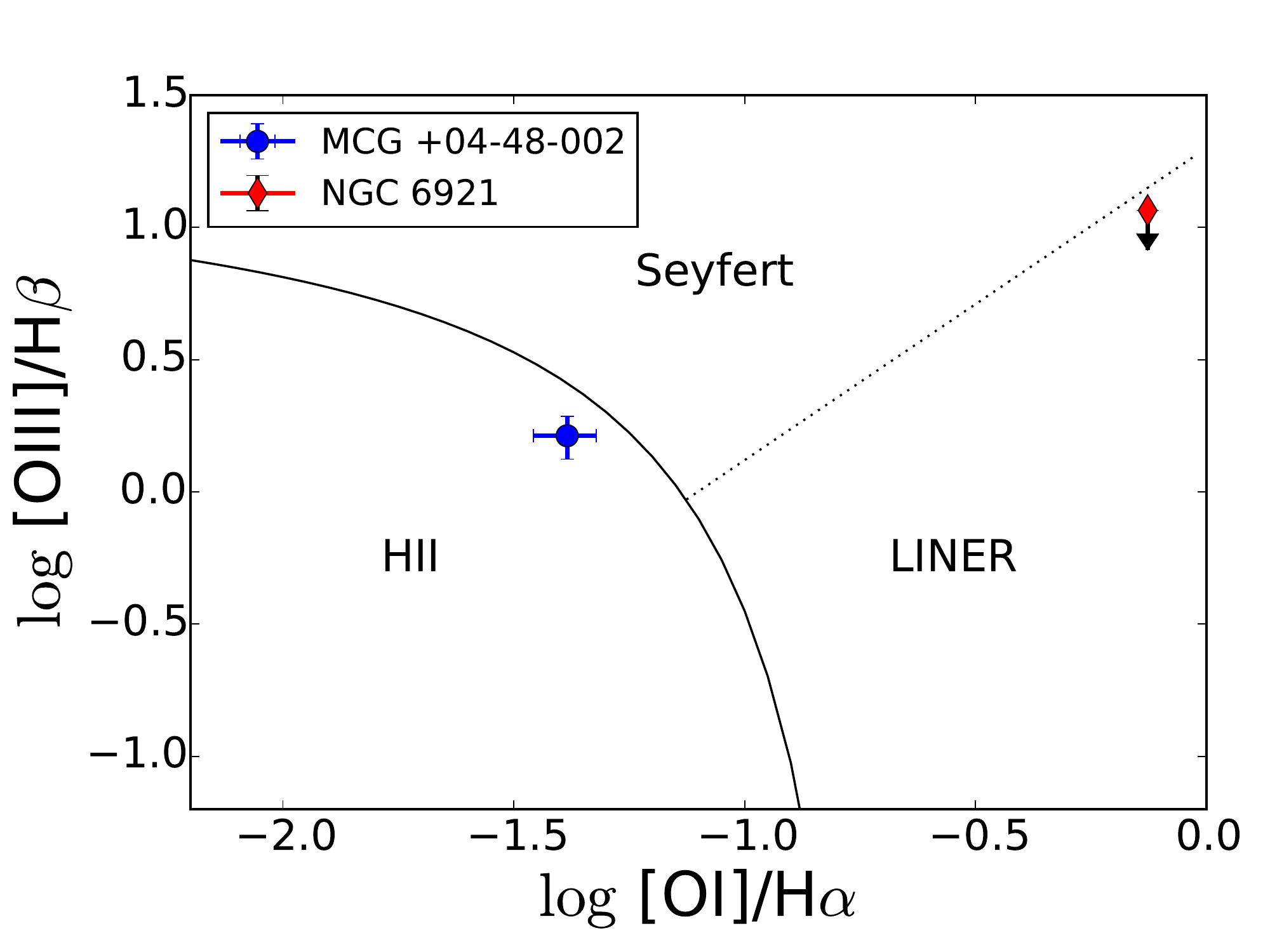}
\includegraphics[width=7cm]{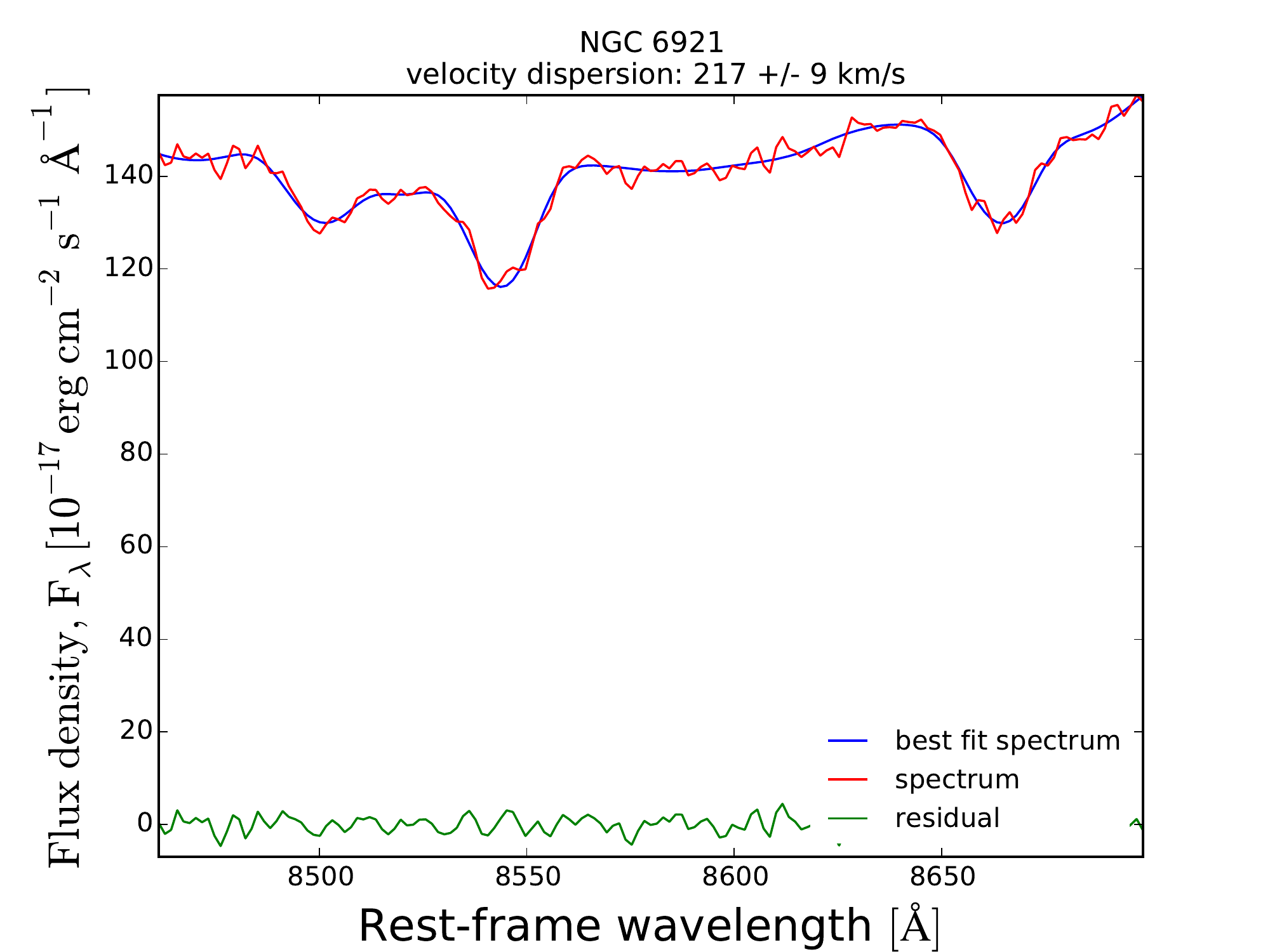}
\includegraphics[width=7cm]{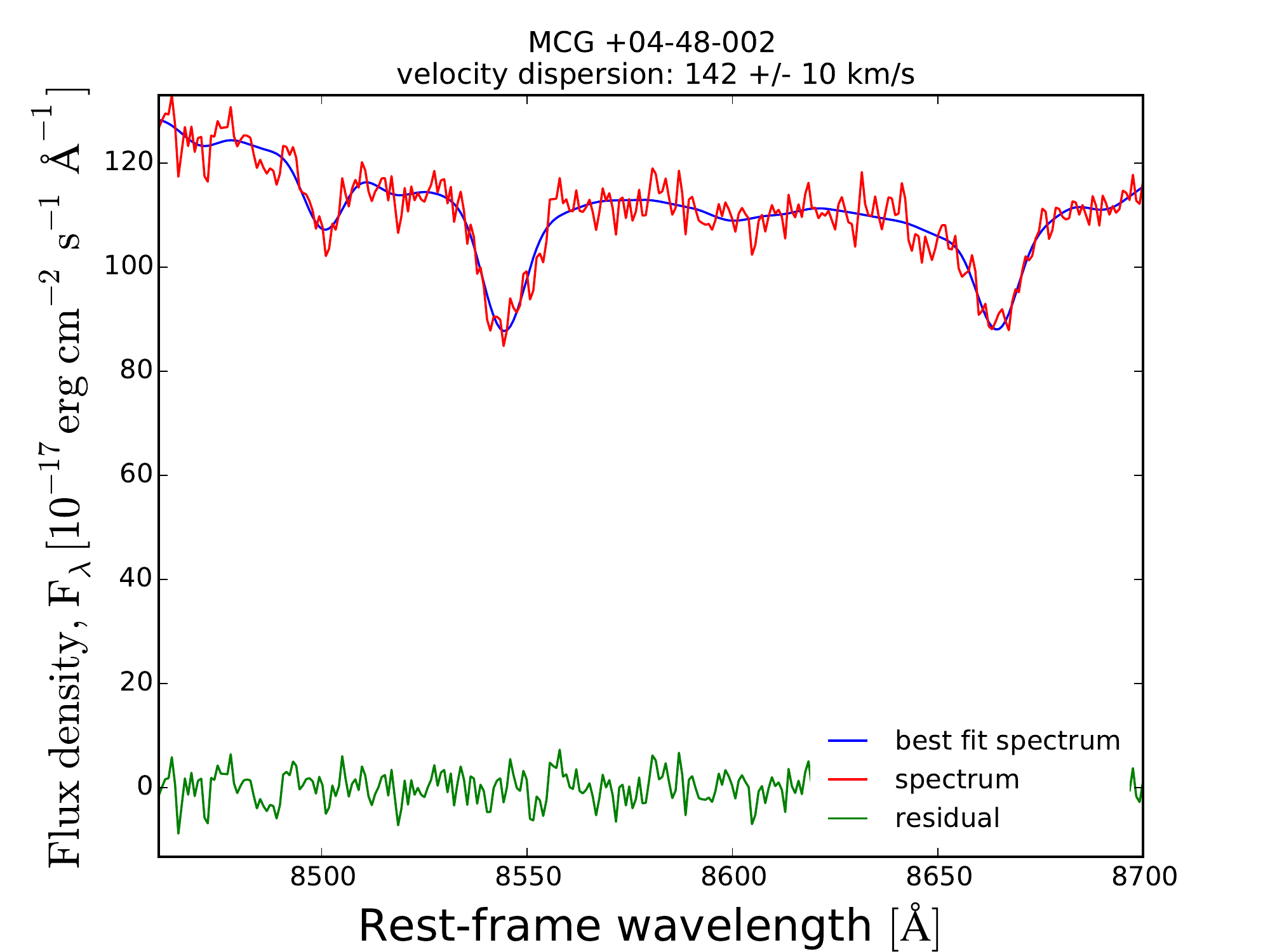}
\caption{Emission and absorption line fits.  {\em Top}: \nii/\Halpha\ (left) and \oi/\Halpha\ (right) diagnostic ratios \citep{Kewley:2006:961}.  The down arrows indicate \Hbeta\ line detection limits. {\em Bottom}: Fits of stellar templates to the Calcium triplet region for \ngc\ (left) and \mcg\ (right).  The data is shown in red, the model in blue, and the residuals are in green.}
\label{bptfig}
\end{figure*}

%54\arcsec 226 and 228  for a 330\,\kmps offset
    We find that \mcg\ is classified as a starburst using the \oi/\ha\ diagnostic and a composite galaxy using the \nii/\ha\ diagnostic \citep[Fig. \ref{bptfig}][]{Kewley:2006:961}.  \mcg\ has strong sky features in the \sii\ region, so this line was not measured.   \ngc\ is classified as a LINER based on the \oi/\ha\ and  \sii/\ha\ diagnostics,  and as an AGN based on \nii/\ha.  For \ngc, the Balmer decrement limit corresponds to $E(B-V)=0.26$.  For \mcg, the Balmer decrement is consistent with no line obscuration (H$\alpha$/H$\beta$=2.62). 
    
    We measure the central velocity dispersion of the Calcium triplet absorption lines to be 217$\pm$9\,km\,s$^{-1}$ for \ngc\ and 142$\pm$10\,km\,s$^{-1}$ for \mcg. Using recent scaling relations from \citet[][]{McConnell:2013:184} these values correspond to $M_{\rm BH}\simeq4\times10^8\,\Msun$ and $M_{\rm BH}\simeq7\times10^7\,\Msun$ for \ngc\ and \mcg, respectively. 
  
%, \ngc\ is classified as a LINER in the \oi/\Halpha\ and  \sii/\Halpha\ diagnostic, respectively, and an AGN using the \nii/\Halpha\ diagnostic

% \mcg\ is classified as a starburst using the \oi/\Halpha\ diagnostic and a composite galaxy using \nii/\Halpha diagnostic. n \ngc, the \Hbeta\ line is undetected and the \Halpha\ line is only weakly detected.
\subsection{X-Ray Variability} \label{sec:xray_variability}
\begin{figure*}	
\includegraphics[width=9.5cm]{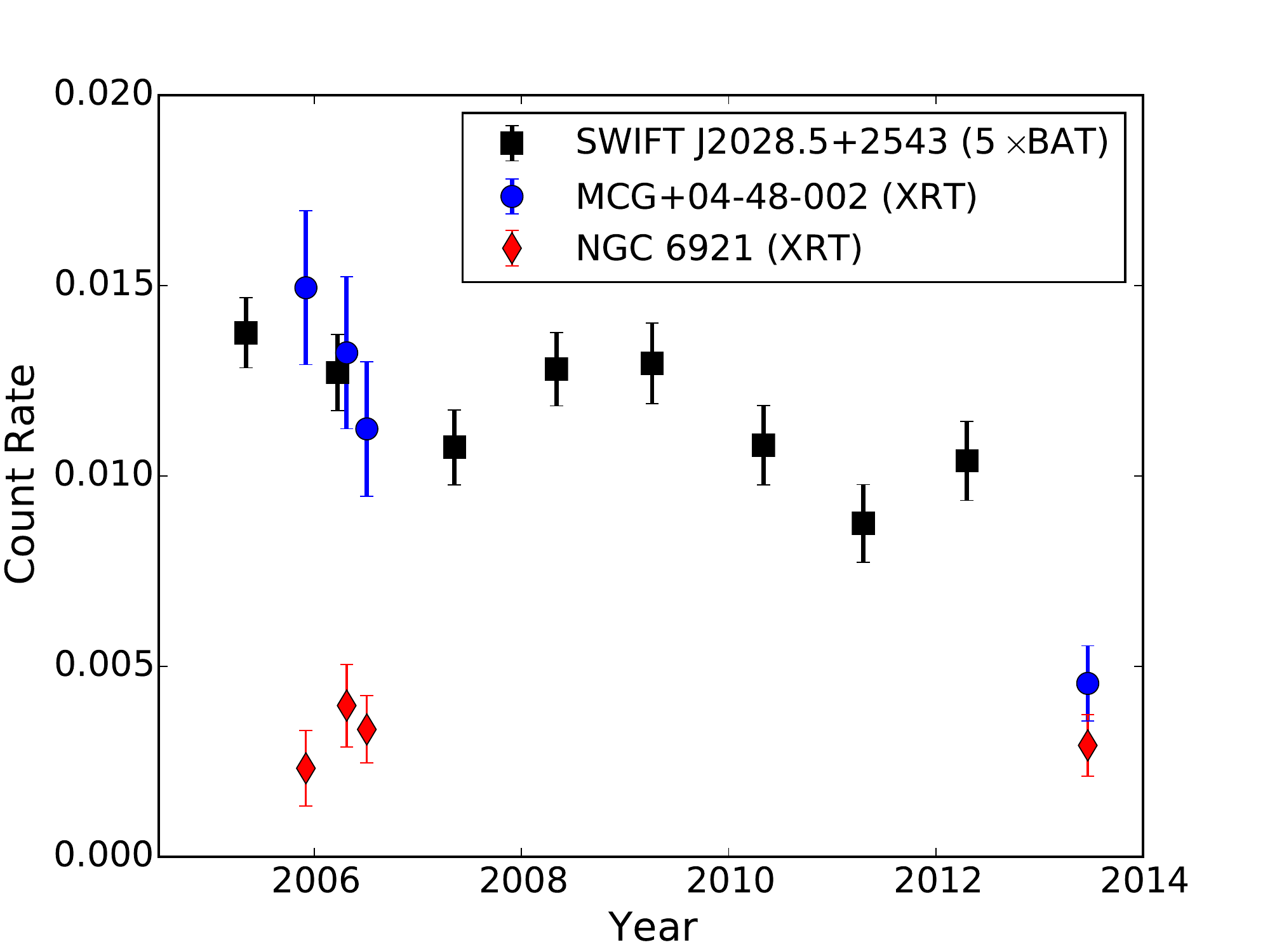}
\includegraphics[width=9.5cm]{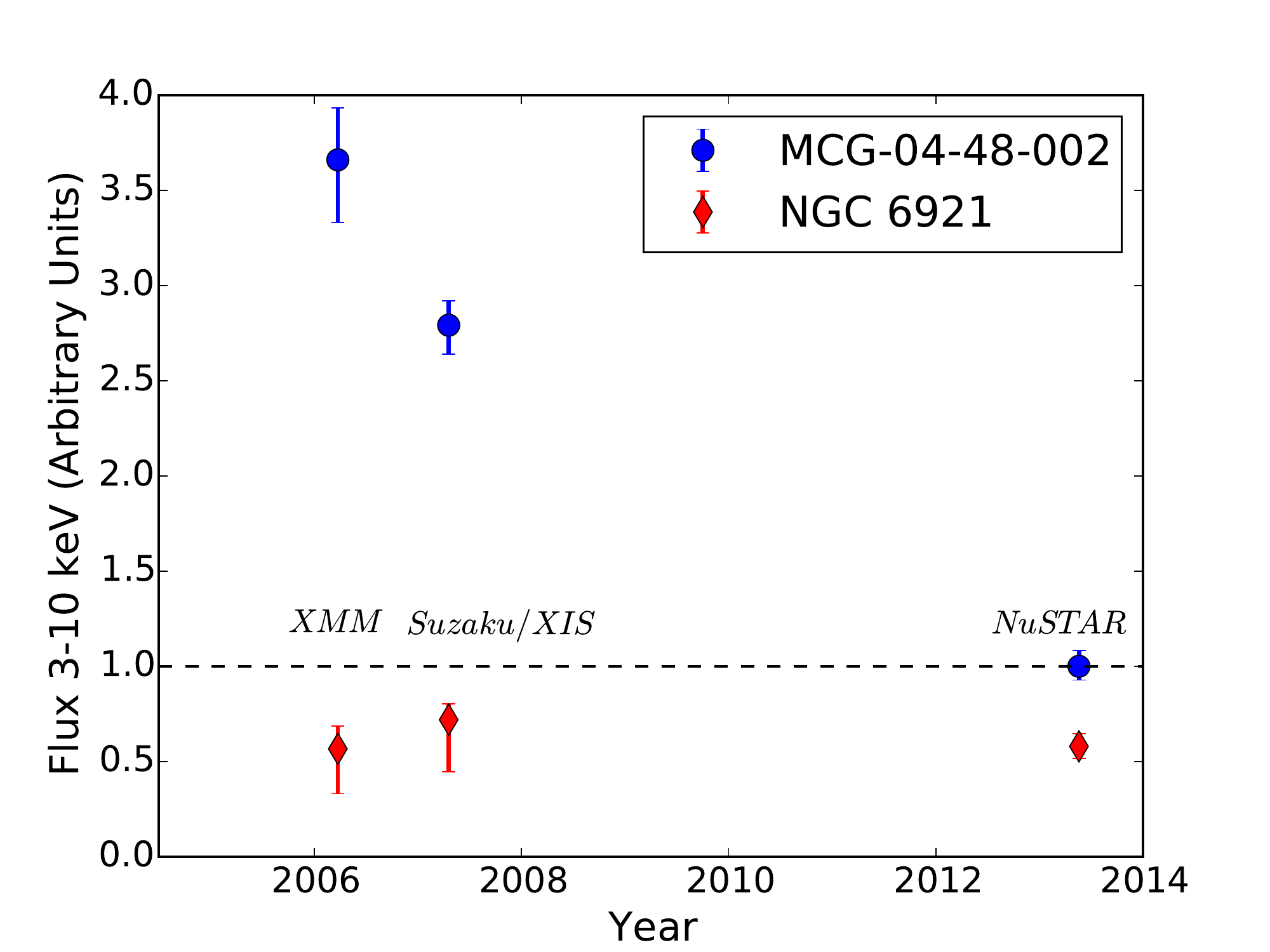}
\caption{{\em Left}: 0.5-10\keV\ count rates of \mcg\ (blue) and \ngc\ (red) from \swiftxrtsh. The BAT lightcurve (black) and associated with emission from both counterparts, binned into year intervals, and is normalized to the Crab.  The \swift BAT count rate has been rescaled by a factor of five for comparison.  {\em Right}: 3-10\,keV flux from \mcg\ and \ngc\ from \xmmsh, \suzakush, and \nustarsh.  The horizontal dashed lines indicates the flux of \mcg\ in the recent \nustar observation.   }
\label{variability_fits}
\end{figure*}

We explore longer term X-ray variability using the \swiftbat 104-month data taken between 2004 and 2013 (Fig.~\ref{variability_fits}).  The spectra of both AGN are blended in \swiftbat because of the low effective angular resolution.  A $\chi^2$ test of the full 14-195\,keV band light curve, binned in one-month intervals, suggests a varying source at the $>99\%$ level.  The BAT light curve shows a significant drop in count rate between the 2005-2009 period (0.0025$\pm$0.0001 Crab) and the 2010-2013 period (0.0020$\pm$0.0001 Crab).  In \swiftxrtsh, for \ngc, there is no count rate variation. In contrast,  \mcg\ varied significantly in the  \swiftxrt count rate between 2013 and earlier observations in 2005 and 2006, when it was higher by a factor of $\simeq$3, at 5$\sigma$ confidence.  \\

%    \swiftxrt also observed the system three times between 2005 Dec~16 and 2006 Jun~3 in addition to the coordinated \nustar observation  on 2013 May~18. 
    We then study the variability in the overlapping 3--10\,keV energy range of \nustarsh, \xmmsh, and \suzakush. We used {\tt Xspec} \citep{Arnaud:1996:17} version 12.8.2 for spectral analysis. We fit the X-ray data using a simple with a power law ($\Gamma$=1.8) plus a normilazation, a Gaussian fixed at 6.4~keV to represent the neutral Fe\,K$\alpha$ line, and obscuration (\NH) for each observation.  For \ngc, we find no evidence of variability, with all the normalizations consistent within uncertainties.  However, for \mcg\ there is significant variability in agreement with the \swiftxrt observations and the \swiftbat data.  We find that the observations from both \xmm and \suzaku data show significantly higher normalizations during 2006-2007, consistent with the higher count rates in \swiftxrt during these times.  There is also no evidence of column density variability in any of the observations.  Further \nustar observations are necessary to study the unresolved high-energy variability seen by \swiftbatsh.
    
    %Finally, we note that \citet{Winter:2008:686} claimed that both sources are highly variable between the \xmm and \suzaku observations. However, from our inspection of the images and the analysis above it is clear that \mcg\ is brighter in the observed 0.5-10\,keV flux than \ngc\ in all X-ray observations.
	
%In summary, our results suggest a dropping flux \mcg\ in 2005-2007 and then a significant drop before the recent 2013 observation \nustar observation. the variability in intrinsic emission or obscuration and the
\subsection{X-ray Spectral Fits} \label{sec:xray_fits}

 We first use a phenomenological model mimicking torus-obscured AGN emission to explore the spectral properties of \ngc\ and \mcg. This model consists of a transmission component, represented by the absorbed power-law model (including Compton scattering), a reprocessed component, represented by the disk-reprocessing model {\tt pexrav} \citep{Magdziarz:1995:837} and to represent the neutral Fe\,K$\alpha$ line emission, a Gaussian fixed at 6.4~keV. We assume a 150-keV high-energy cutoff, typical of Seyfert nuclei \citep{Fabian:2015:4375}. We also include a scattered power-law component with photon index equal to that of the intrinsic spectrum and relative normalization of $\sim1$\%, as expected from Thomson-scattering by the free electrons outside of the putative AGN torus. Due to the limited photon statistics, we use Cash statistics, although we also report $\chi^2$ values due to their straightforward interpretability. 
 
 %All fits discussed below have null-hypothesis probabilities exceeding 20\%. that obscures most of the intrinsic continuumfor fitting
 
 For \ngc, the best fit of the \nustar and \swiftxrt data ($\chi^2/{\rm d.o.f.}=36/43$) is with $\Gamma=1.8_{-0.3}^{+0.2}$ and \NH$=\left(1.9\pm0.5\right)\times10^{24}$\,cm$^{-2}$, which corresponds to a Compton-thick scenario with the reprocessed component contributing $\sim10$\% of the total 10--50~keV flux.  This model implies an intrinsic 10--50\,keV luminosity of $1.9\times10^{43}$\,erg\,s$^{-1}$.  An alternative solution with slightly higher $\chi^2$ ($\chi^2/{\rm d.o.f.}$=47/44) is also found by allowing the reprocessed continuum to be absorbed. This implies a significantly higher column density ($\gtrsim5\times10^{24}$\,cm$^{-2}$) and a higher intrinsic luminosity \citep[as in e.g.,]{Balokovic:2014:111,Brightman:2015:41}.  Equivalent widths of the neutral Fe\,K$\alpha$ line at 6.4\,keV range between 0.5--2.4\,keV, as expected from a Compton-thick torus.

 %Since \ngc\ is faint below 10\,keV, we have to fix the photon index to achieve reasonable spectral solutions using the in \xmm and \suzaku observations alone. Fitting the \suzaku data with fixed $\Gamma=1.9$, a best fit ($\chi^2/{\rm d.o.f.}=9.6/9$) is achieved with borderline Compton-thick column density, $\left(1.2_{-0.4}^{+0.6}\right)\times10^{24}$\,cm$^{-2}$. For a softer assumed $\Gamma$, \NH moves deeper into the Compton-thick regime, while the opposite is true for a harder $\Gamma$. Results are similar for the \xmm data: $\chi^2/{\rm d.o.f.}=10.9/9$ and \NH$=\left(8_{-3}^{+5}\right)\times10^{23}$\,cm$^{-2}$.      

 %The intrinsic 10--50~keV luminosity based on the best-fit models are $1.1\times10^{43}$\,erg\,s$^{-1}$ for the \suzaku observation and $2.9\times10^{42}$\,erg\,s$^{-1}$ for the \xmm observation.  

The simultaneous \nustar and \swiftxrt spectra of \mcg\ are fitted well with the model described above ($\chi^2/{\rm d.o.f.}=57/49$) for $\Gamma=1.8\pm0.2$ and \NH$=\left(1.0\pm0.3\right)\times10^{24}$\,cm$^{-2}$. In this solution, the relative normalization of the scattered continuum is 0.4\% (<2\% with 90\% confidence) and the reprocessed continuum contributes 20\% of the observed 10--50~keV flux (2--130\% within the 90\% confidence interval). The intrinsic 10--50~keV luminosity based on this model is $6.5\times10^{42}$\,erg\,s$^{-1}$. Equivalent widths of the neutral Fe\,K$\alpha$ line at 6.4\,keV are  0.3--1.2\,keV.    If we include the archival soft X-ray data assuming a normilization offset because the source was significantly brighter, the $N_{\rm H}=\left(8\pm3\right)\times10^{23}$\,cm$^{-2}$, suggesting that the column density did not change drastically between observations though the lack of $>$10 keV coverage in the earlier observation limits our constraints.

\begin{figure*}	
\includegraphics[width=9.5cm]{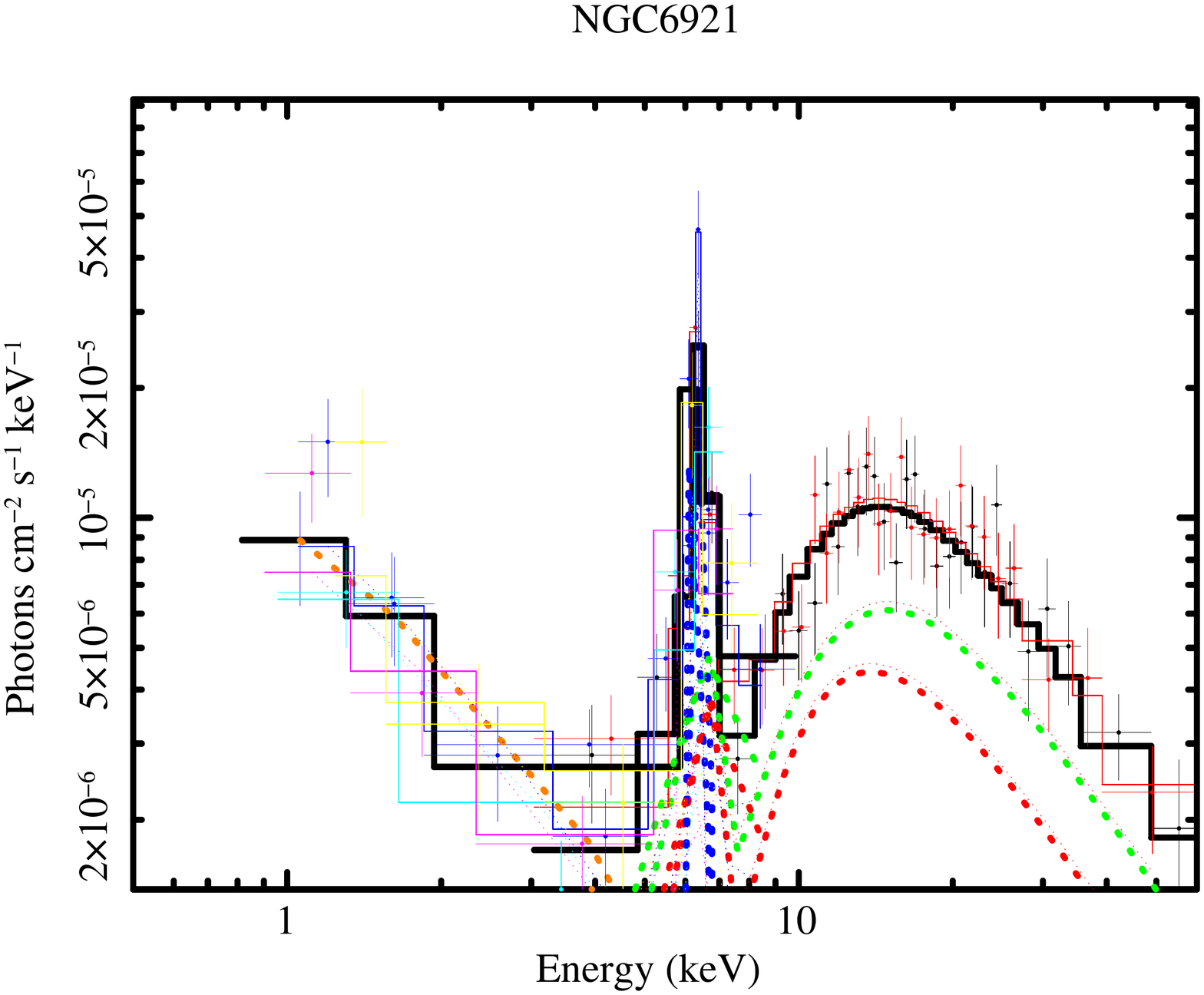}
\includegraphics[width=9.5cm]{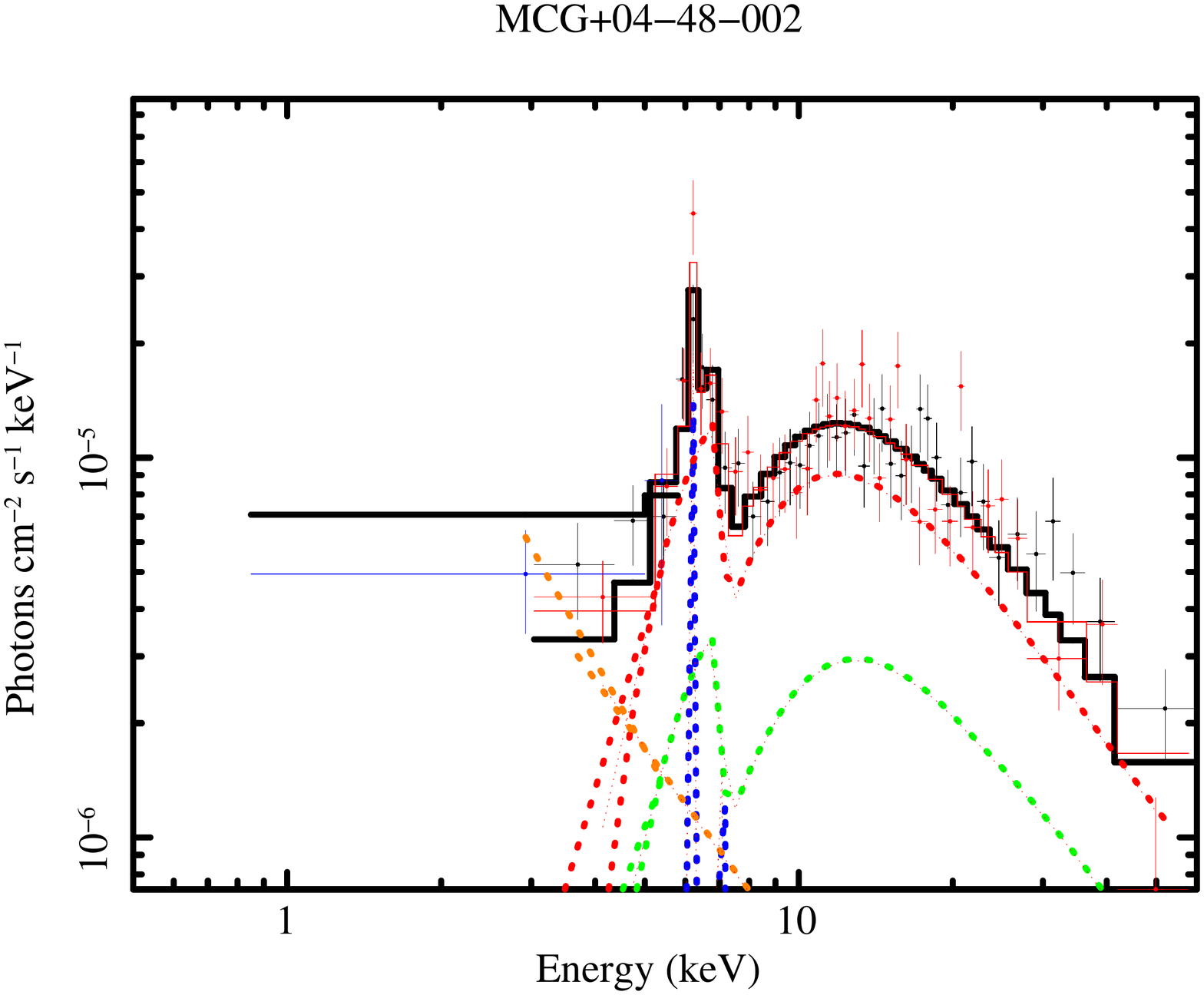}
\caption{X-ray spectra of \ngc\ (left) and \mcg\ (right) and best-fit {\tt MYtorus}-based model shown binned to match the unfolded data.  \nustar is shown in black (FPMA) and red (FPMB).  The soft X-ray data are from \xmm (PN-blue and MOS-yellow), \suzaku (magenta), and \swiftxrt (cyan) for \ngc.  Since we found significant variability in \mcg, we only use the simultaneous \swiftxrt observation in the soft X-rays (blue).  The sum of the model is represented by a solid black line.  The model components are represented by dashed lines indicating the zeroth-order transmitted continuum through photoelectric absorption ($\mathtt{MYTZ}$, red), the scattered/reflected component ($\mathtt{MYTS}$, green), and fluorescent emission-line spectrum ($\mathtt{MYTL}$, dark blue).  At softer energies ($<$3~keV), there is a model component for scattered AGN emission on larger scales in the host galaxy ($f_{\rm scatt}$, orange).  }
\label{modelM_fits}
\end{figure*}

Following the strategies of past studies of single AGN observed with \nustar \citep[e.g., ][]{Balokovic:2014:111,Gandhi:2014:117,Koss:2015:149}, we fit the X-ray spectra with the \mytorus model \citep[]{Murphy:2009:1549} shown in Fig.~\ref{modelM_fits}.  For \ngc\, we combine the \suzakush, \xmmsh, and \swiftxrt data, with an offset of 10\% variability allowed for telescope cross-normalization.  We find $\Gamma$=1.7$\pm$0.1, $N_{\rm H}=1.4\pm0.1\times10^{24}$ \nhunit, and $\theta_{\rm inc}=89\degree^{+1}_{-8}$, consistent with a Compton-thick torus observed nearly edge-on ($\chi^2/{\rm d.o.f.}=67/79$).  Using this model we find $L_{\rm \ 2-10 \ keV}^{\rm obs/int}= (0.3/4.4)\times 10^{42}$~\ergpersec\ and $L_{\rm 10-50 \ keV}^{\rm obs/int}= (3.6/5.5) \times 10^{42}$~\ergpersec\ for \ngc.  For \mcg, we limit our fit to the simultaneous \nustar and \swiftxrt spectra because of variability.  Using the \mytorus model with fixed $\Gamma$=1.9, $\theta_{\rm inc}=85\degree$, and $\theta_{\rm tor}=60\degree$, we obtain a column density of \NH$=1.0^{+0.4}_{-0.2}\times10^{24}$\, \nhunit, in agreement with our simpler model ($\chi^2/{\rm d.o.f.}=64/83$).  We find $L_{\rm \ 2-10 \ keV}^{\rm obs/int}= (0.2/3.7) \times 10^{42}$~\ergpersec\ and $L_{\rm 10-50 \ keV}^{\rm obs/int}= (3.2/6.1) \times 10^{42}$~\ergpersec\ with this model.  

%In summary, we conclude that \ngc\ should be considered a bona-fide Compton-thick AGN, while \mcg\ is heavily obscured to Compton-thick.  Deeper observations with \xmm and \nustar are needed for tighter constraints on the column density and variability in \mcg.

%to the \swiftxrt measurements

\subsection{Intrinsic Luminosity and Eddington Ratio} \label{sec:intrin_lum}

    One estimate of the intrinsic AGN luminosity comes from \oiii.  The \oiii\ luminosity is $1.1\times10^{39} \ergpssh$ for \ngc\ and $1.5\times10^{39} \ergpssh$ for \mcg.  Based on the relation from a study of 351 BAT AGN \citep[][]{Berney:2015:3622}, we expect $L_{\oiii}\simeq4.9\times10^{40} \ergpssh$ for \ngc\ and $L_{\oiii}\simeq4.0\times10^{40} \ergpssh$ for \mcg\ as inferred from the 2-10\,keV intrinsic luminosity derived from the X-ray spectra.  This values are 25--45 times higher than observed, implying both sources have weak \oiii\ emission, though some extended \oiii\ is likely missed because the slit widths correspond to $\approx$150\,pc and $\approx$450\,pc for \ngc\ and \mcg, respectively.  

Another estimate of intrinsic AGN luminosity is $L_{12\, \micron}$ which is measured using the photometry from the {\em Wide-field Infrared Survey Explorer final catalog release} at $7.0\times10^{42} \ergpssh$ for \ngc, and $3.0\times10^{43} \ergpssh$ for \mcg.  The expected unabsorbed 2-10\,keV luminosity, based upon the mid-IR/X-ray correlation is then $\simeq1.5\times10^{43} \ergpssh$ in \ngc\ and $\simeq6.2\times10^{43} \ergpssh$ in \mcg\ \citep{Gandhi:2009:457,Asmus:2015:766}.  The estimate of intrinsic AGN luminosity of \mcg\ from $L_{12\, \micron}$ is then more than a factor of 3 higher than our X-ray measurement, suggesting we underestimated the intrinsic luminosity in the X-rays and \mcg\ may be heavily Compton-thick.   We note, however, that some of the IR emission may be from star formation from \mcg\ being a luminous infrared galaxy \citep[LIRG,][]{Armus:2009:559}. 

We use a bolometric correction of 15 \citep{Vasudevan:2009:1124} to convert the unabsorbed 2-10\,keV luminosities to bolometric luminosities.  This implies a bolometric luminosity of $\simeq7\times10^{43} \ergpssh$ for \ngc\ and $6\times10^{43} \ergpssh$ for \mcg.  Combined with the measured SMBH mass we estimate the Eddington fraction, $L_{\rm Bol}/L_{\rm Edd}$, where $L_{\rm Edd}$ is the Eddington luminosity.  The Eddington ratio is then $\approx$0.001 for \ngc\ and $\approx$0.009 for \mcg.

\section{Discussion}  

We have discovered heavy obscuration in the dual AGN associated with the \swiftbat source SWIFT\,J2028.5$+$2543 using \nustarsh.  \ngc\ is obscured by a Compton-thick column which is well constrained by the \nustar data, but only poorly constrained by the archival soft X-ray data.  \mcg\ is obscured by heavy to Compton-thick material along the line of sight ($N_{\rm H}\approx 1\times10^{24}$\,cm$^{-2}$), with deeper observations required to better understand the variability.  Both sources are severely diminished in the 2-10 keV band ($L_{\rm \ 2-10 \ keV}^{\rm obs/int}<0.1$) while the majority of the $>10$ keV emission is detected, illustrating the importance of \nustar and \swiftbatsh. On average, the two AGN are similarly luminous (within a factor of $\simeq2$).  We note that in \citet{Winter:2008:686} an error was likely made in identifying the two sources in the \xmm\ image, such that their names were switched. This led \citet{Winter:2009:1322} to claim that NGC 6921 was highly variable between the \xmm and \suzaku observations. However, from our results it is clear that \mcg\ is brighter in the observed 0.5-10\,keV emission than \ngc\ in all X-ray observations and \ngc\ shows no significant evidence of variability between observations.  

%The estimated flux in the \swiftbat band, based on \nustarsh, are significantly above the limiting sensitivity of $1.3\times10^{-11}$\,erg\,s$^{-1}$\,cm$^{-2}$ which \swiftbat achieves over 90\% of the sky.  and have similar observed count rates in the \swiftbat band (within 5\%)

Despite being bright nearby X-ray selected AGN, these sources would be missed in large optical spectroscopic AGN catalogs \citep[e.g.,][]{Kauffmann:2003:1055}.   For \ngc, high levels of dust extinction likely contribute to the \hb\ non-detection, and for the LIRG \mcg, the intense star formation may overwhelm the AGN photoionization signature.  This is typical of about 5\% of BAT-selected AGN \citep[][]{Smith:2014:112,Schawinski:2015:2517} and is more common to BAT-selected AGN in ongoing mergers which tend to have lower \oiii/X-ray ratios \citep{Koss:2010:L125}.

%$1.9\times10^{-11}$\,erg\,s$^{-1}$\,cm$^{-2}$ and $2.3\times10^{-11}$\,erg\,s$^{-1}$\,cm$^{-2}$,

The pair shows spectroscopic signatures typical of merger-triggered dual AGN rather than a chance association. The small line of sight velocity offset ($\simeq$140\,km\,s$^{-1}$), is typical of dual AGN found using other techniques \citep[50--300\,km\,s$^{-1}$,][]{Comerford:2013:64}.  AGN bright enough to be detected by \swiftbat are rare \citep[e.g., 0.02 per square degree on the sky,][]{Baumgartner:2013:19}.  Since there are only three other nearby ($\pm$200\,km\,s$^{-1}$) BAT sources of \mcg\ in the entire sky, the chance possibility of a random BAT AGN at the same redshift within 91\arcsec\ is very small ($<10^{-8}$).    Since the system is near the Galactic plane ($\delta_{\rm Gal}=-7^{\circ}$), large foreground Galactic extinction ($\approx$1 mag) make optical detection of merger features like tidal tails difficult.  Mapping the distribution and line-of-sight velocity of the atomic gas in the 21-cm line of neutral hydrogen to search for gas-rich material thrown off in such encounters \citep[e.g.,][]{Hibbard:1996:655} would be helpful to study the merger. However, no sufficiently high resolution maps currently exist from all sky surveys for this sky region.   

The heavily obscured dual AGN in \mcg\ and \ngc\ share several properties with the BAT-detected Compton-thick dual AGN NGC 6240.  Both systems are Luminous Infrared Galaxies (LIRGs) which are rare in the nearby universe ($z<$0.03) and in the BAT sample \citep{Koss:2013:L26}.  The intrinsic 2-10\,keV luminosities of the dual AGN are nearly equal, which is similar to NGC 6240 \citep{Puccetti:2016:A157}, but not common in typical BAT-detected dual AGN  where the median ratio is 11 \citep[][]{Koss:2012:L22}.  NGC 6240 is also classified as a LINER, similar to NGC 6921, which is found in only $\approx$4\% of the BAT sample (Koss et al., in prep).  NGC 6240, however, is at a 1.4\,kpc separation,  whereas this system has a larger 25.3\,kpc separation, suggesting even the early merger phase (20-30 kpc) can contribute to both AGN's obscuration.  Larger statistical studies to understand merger-triggered obscuration with \nustar are currently being performed  in BAT AGN (Koss et al., in prep) and in LIRGs (Ricci et al., in prep).

%of merging AGN

%Based on simulations and using the ratio of the stellar masses (1:3) and spatial separation separation the merger would take 250\,Myr for final coalescence \citep{VanWassenhove:2012:L7}.  
%Dual AGN are common in the BAT sample \citep[8\%, ][]{Koss:2012:L22} and among gas rich major mergers at small separations the rate is 100\% (13/13), suggesting this system is very likely to host a dual AGN.  
\acknowledgments
We acknowledge the:  Ambizione fellowship grant PZ00P2\textunderscore154799/1 (M.K.) and the Joanna Wall Muir and the Caltech Student Faculty Program (A.G.).     This work was supported under NASA Contract No.\ NNG08FD60C, and made use of \nustar mission data, a project led by the California Institute of Technology, and managed by the Jet Propulsion Laboratory.

\clearpage

{\it Facilities:} \facility{Swift},  \facility{KPNO:2.1m}, \facility{Hale}, \facility{NuSTAR},  \facility{XMM},  \facility{Suzaku}

\bibliographystyle{apj}
%\bibliography{/Applications/astronat/bibfinal}

\end{document}